\documentclass[amsmath,amssymb,epsfig,showpacs,preprint]{revtex4}

\usepackage{graphicx}
\usepackage{dcolumn}
\usepackage{bm}
\newcommand{\abs}[1]{\left\vert#1\right\vert}

\begin{document}

\title{Interactions of Parametrically Driven Dark Solitons. I: \\
N\'eel-N\'eel and Bloch-Bloch interactions}

\author{I.V. Barashenkov}
 \email{Igor.Barashenkov@uct.ac.za; igor@odette.mth.uct.ac.za}
\affiliation{Department of Physics, University of Bayreuth, D-95440
Bayreuth, Germany}
\altaffiliation{On sabbatical leave from 
University of Cape
Town. Permanent address: Department of Applied Mathematics,
University of Cape
Town, Rondebosch 7701, South Africa}
\author{S.R. Woodford}
 \email{s.woodford@fz-juelich.de}
\affiliation{Theorie I, Institut f\"ur Festk\"orperforschung,
Forschungszentrum J\"ulich,  D-52428 J\"ulich, Germany} 
\author{E.V. Zemlyanaya}
 \email{elena@jinr.ru}
\affiliation{Joint Institute for
Nuclear Research, Dubna 141980, Russia} 

\date{\today}

\begin{abstract}
We study interactions between the dark solitons of 
the parametrically driven nonlinear Schr\"odinger equation, 
Eq.(\ref{NLS}).
When the driving strength, $h$, is below  $\sqrt{\gamma^2 +1/9}$,
two well-separated  N\'eel walls may repel or attract.
They repel if their initial separation $2z(0)$ is larger than
the distance $2z_u$ between the constituents in the unstable
stationary complex of two walls. They  
attract and annihilate if $2z(0)$ is smaller than $2z_u$.
Two N\'eel walls with $h$ lying between $\sqrt{\gamma^2 + 1/9}$
and a threshold driving
strength $h_{sn}$ attract for $2z(0)<2z_u$ and
evolve into
a stable stationary bound state for $2z(0)>2z_u$.
Finally, the N\'eel walls with $h$ greater than  $h_{sn}$ attract and
annihilate --- irrespective of their
initial separation.  
Two Bloch walls of opposite chiralities attract, while Bloch walls of like
chiralities repel --- except 
near the critical driving strength,
where the difference between the like-handed and 
oppositely-handed walls becomes negligible. 
In this limit, similarly-handed walls 
at large separations repel while those placed at shorter
distances may start moving in the same 
direction or transmute into an oppositely-handed pair and attract.
The collision of two Bloch
walls or two nondissipative N\'eel walls typically produces a quiescent or
moving breather. 
\end{abstract}


\maketitle

\section{Introduction}
\label{Intro}

This paper deals with the parametrically driven, repulsive nonlinear Schr\"odinger 
(NLS) equation:
\begin{equation}
\label{NLS}
i\partial_t \Psi + {\textstyle \frac 12} \partial_{X}^2 \Psi 
+ \Psi - |\Psi|^2\Psi = h\Psi^* - i\gamma\Psi.
\end{equation}
Here  
$h$\ is the strength of the parametric driving and $\gamma$\ is the damping
coefficient. In the absence of damping, i.e. when $\gamma=0$, this equation 
has the same stationary Bloch- and N\'eel-wall solutions as the 
(2:1)-resonantly forced
Ginzburg-Landau equation and the relativistic Montonen-Sarker-Trullinger-Bishop 
model \cite{Sarker, Montonen, Niez, Niez2}. However,
unlike the  
Ginzburg-Landau and the relativistic model, where only one of these solutions (the Bloch wall) is linearly 
stable in their region of
coexistence, the NLS equation exhibits Bloch-N\'eel bistability. Furthermore, both domain-wall solutions 
of the NLS (also known as dark solitons or kinks) 
can move with constant velocity, 
and the moving walls also turn out to be linearly stable \cite{OurPaper}. 
This multistability is a rare phenomenon
and leads one to wonder about the outcome of the soliton-soliton interaction. 
The problem that is of ultimate interest to a physicist, is formulated as follows: 
Given an initial configuration of many different
walls, will the asymptotic solution, as $t \rightarrow \infty$, consist of predominantly 
Bloch walls, N\'eel walls or possibly some 
other, more complicated, structures? 
Will the walls form widely spaced, equidistant lattices, or tightly bound clusters? 
In this paper we make the first step towards answering these questions.

 When $\gamma \neq 0$,  equation (\ref{NLS}) 
 still has a dark soliton solution 
(in the form of the N\'eel wall); this solution was shown
to be stable for all $\gamma$\ \cite{OurPaper}. Our analysis of soliton interactions 
will naturally include the case of nonzero damping, 
especially in view of
the wide range of applications of Eq.(\ref{NLS}). 

Equation (\ref{NLS}) was indeed derived in a broad variety of physical situations.
In fluid dynamics, the repulsive (``defocusing'') parametrically driven NLS
describes the amplitude of
the water surface in a vibrated channel with large width-to-depth 
ratio \cite{Elphick_Meron,Larraza,Chen}. (The case of the small width-to-depth ratio gives rise to the attractive NLS 
\cite{Elphick_Meron,Larraza,Chen}.)
The same equation arises as an
amplitude
equation for the upper cutoff mode 
in a chain of parametrically driven, damped nonlinear oscillators.
 In the optical context, it was derived for the
doubly resonant $\chi^{(2)}$
optical parametric oscillator in the limit
of large second-harmonic detuning \cite{Trillo}. 
Next, stationary solutions of Eq.(\ref{NLS}) with $\gamma = 0$\ minimise the 
Ginzburg-Landau 
free energy of the anisotropic $XY$-model. Here $F=\int {\cal F} d {X}$, 
where
\[
{\cal F} = 
{\textstyle \frac12} (\partial_X {\bf M})^2 -(1+h){\bf M}^2 + {\textstyle
\frac12} {\bf M}^4  +
2 h M_y^2 + {\cal F}_0,
\]
and ${\bf M}(X)=(0,M_y,M_z)$ is the magnetisation vector 
whose nontrivial components serve as the real and imaginary parts
of the complex field $\Psi(X)$ in (\ref{NLS}): 
$\Psi=M_y+i M_z$.
This model appeared in studies of stationary domain walls in easy-axis ferromagnets
near the Curie point \cite{XY}. 
 {\it Nonstationary\/} 
magnetisation configurations
were considered in the overdamped limit: $\Psi_t=
-\delta F/\delta  \Psi^*$ \cite{Coullet}.
The damped hamiltonian dynamics $\Psi_t=-i\delta F/\delta \Psi^*
-  \gamma \Psi$ provides a sensible alternative; this is precisely
our equation (\ref{NLS}).

Finally we note that Eq.(\ref{NLS}) also 
arises in a completely different magnetic context --- that of a weakly anisotropic easy-plane
ferromagnet in a constant external magnetic field \cite{OurPaper}. 
Here, the external field (chosen parallel to the $z$-axis) 
forces the magnetisation to be
almost homogeneous, and Eq.(\ref{NLS}) describes small deviations 
$\epsilon \Psi = M_x + i M_y$\ 
from the uniform  magnetisation ${\bf M} \approx (0,0,1)$. (Here $\epsilon$\ is a small
parameter.)  In the $XY$-model, on the other hand, Eq.(\ref{NLS})
governs the magnetisation vector itself.

The literature devoted to the interactions of dark solitons and kinks is 
quite extensive (although perhaps not as vast as for bright solitons
and pulses). Kinks are known to attract antikinks in real-valued, single-component 
Klein-Gordon equations, such as the sine-Gordon and 
$\phi^4$-theory \cite{RajKink}, with and without damping terms \cite{KawOhta}. 
The same is true for kinks of the real Ginzburg-Landau
equations $u_t = u_{xx} - V'(u)$\ \cite{KawOhta}. These results admit simple interpretations; 
the kink-antikink pair converges in
order to minimise its total energy in the former case, 
and to minimise the value of its Lyapunov functional in the latter. Next, the dark
solitons of the (undriven) NLS equations are known to repel \cite{ZakShab,KivsharReview}. 
This can also be understood as an attempt to minimise the
energy of the pair, at the expense of the gradient of the phase. 
Proceeding to the domain-wall solutions of the parametrically driven
Ginzburg-Landau equation
\begin{equation}
\label{CGLE}
\partial_t \Psi = {\textstyle \frac 12} \partial_{X}^2 \Psi + \Psi - |\Psi|^2\Psi - h\Psi^*, 
\end{equation}
the Bloch wall and antiwall of opposite chiralities were shown to attract, while those of like chiralities repel \cite{Korz,Tutu}. 
(Here $h <\frac 13$, the region of stability of Bloch walls.) The N\'eel wall and antiwall always repel in their stability region 
($h > \frac 13$) \cite{Korz}.
Unlike the case without driving, the interpretation of these interactions is not straight-forward though.

As for the driven NLS equation, 
the behaviour of its Bloch and N\'eel walls is even 
less predictable at the intuitive level, while
the variety of possible interacting partners is much wider due to the 
multistability.
 We will see
that the interaction pattern is quite complicated indeed.
The character of interaction (repulsion vs attraction) 
will depend both on the driving strength and the interwall
separation. We will also show that it is influenced by stationary
complexes of walls, stable and unstable.

This paper constitutes the first part of our project;
here, we restrict ourselves to interactions of the walls
of the same type, i.e. N\'eel-N\'eel and Bloch-Bloch interactions.
The analysis of the nonsymmetric situations, i.e. N\'eel-Bloch 
interactions, requires a different mathematical formalism 
and will be presented separately. (See the following publication
\cite{BW2}.)

The dynamics of the {\it repelling\/} solitons is, in a sense, trivial:
 if the walls are initially at rest, they will simply diverge to the 
infinities. Less obvious is what the  collision of two
{\it attracting\/} walls will result in.
The study of the asymptotic (as $t \to \infty$) attractors 
arising in the parametrically driven NLS constitutes the
second objective of our work.
We will show that if the dynamics are
dissipative, then, depending on the strength of the driving,
colliding walls either annihilate or form a stable stationary bound state. 
In contrast to this, undamped collisions will be found to 
always produce a breather, a spatially localised, temporally oscillating
structure. Depending on the initial conditions, the breather
propagates or remains motionless, and in either case
is found to persist indefinitely. 

The outline of this paper is as follows. 
The two fundamental solutions of Eq.(\ref{NLS}), the Bloch  and N\'eel wall,
are introduced in the 
next section.
For the analysis of the  Bloch-Bloch and
N\'eel-N\'eel interaction we use the variational method, 
under the assumption of well-separated walls.
The method is detailed in section \ref{Variational} and 
the resulting finite-dimensional systems 
are analysed in sections 
 \ref{NeelVar}, \ref{OppVar}, and \ref{LikeVar}. 
 Section \ref{NeelVar} deals with the N\'eel
 walls; section \ref{OppVar} is devoted to the Bloch walls of
 the opposite chirality, while
 the like-chirality walls are examined in section \ref{LikeVar}.
 The conclusions of the variational analysis 
 have been verified in direct numerical simulations of the 
 full partial differential equation. The  numerical simulations allow 
   to advance beyond the limit of well-separated walls; 
 in particular, we use this approach to examine 
the outcome of soliton collisions.
We have allocated  a separate section (section \ref{Simulations})
 to the N\'eel-N\'eel simulations while the Bloch-Bloch simulations
 are reported in  sections \ref{OppVar} and \ref{LikeVar},
 along with the corresponding variational results.
The  nontrivial attractors mentioned above --- 
the stationary bound state in the case of dissipative dynamics, 
and the breather in the undamped
system --- are further investigated in 
sections \ref{DBS} and \ref{Breathers}, respectively. Finally, the main
results of this work are summarised  in section \ref{Conclusions}.

\section{Dark solitons: preliminaries}

Dark solitons  are
localised patches of low intensity ($|\Psi|^2 \ll 1$) 
in a high intensity background ($|\Psi|^2 = \mathcal{O}(1)$). The admissible backgrounds are described by the 
stationary, spatially homogeneous nonzero
solutions of Eq.(\ref{NLS}),
\begin{equation}
\Psi^{(\pm)}_{\mbox{flat}} = iA_{\pm} e^{i\theta_{\pm}},
\label{psiflat}
\end{equation}
\noindent where 
\begin{eqnarray}
A_{\pm} = \sqrt{1\pm \sqrt{h^2 - \gamma^2}}, 
\nonumber \\  2\theta_{+}= \arcsin{\frac{\gamma}{h}}, \quad 2\theta_{-}= \pi -
\arcsin{\frac{\gamma}{h}}.  
\label{star}
\end{eqnarray}
As one can easily check, for $h \leq \gamma$\ the zero solution is the only stable background and so dark solitons can
only exist for 
$h > \gamma$;
this is the condition we are implicitly assuming in this paper. 
It can be shown that $\Psi^{(+)}_{\mbox{flat}}$\ is stable
 for all values of $h>\gamma$\ while 
$\Psi^{(-)}_{\mbox{flat}}$\ is always unstable.
We will only consider dark solitons existing over
the stable background so that all  solutions obey $|\Psi|^2 \rightarrow A_+^2$\ as $|X|\rightarrow \infty$. Accordingly, $A$\ stands for
$A_+$\ and $\theta$\ for $\theta_+$\ for the rest of the paper.  

It will be convenient to transform
 Eq.(\ref{NLS}) so that the asymptotic solution is 
 independent of $h$\ and $\gamma$.  To  this end, 
we set
\begin{equation}
\Psi(X,t) = iA e^{i\theta}\psi(x,t), \quad x = AX.
\label{MsEllen}
\end{equation}
(We have also rescaled the spatial dependence for later convenience.)
Under this transformation, Eq.(\ref{NLS}) becomes
\begin{equation}
i \psi_t + {\textstyle \frac {A^2}{2}} \psi_{xx} - A^2 |\psi|^2 \psi + \psi 
 + (A^2 - 1)\psi^* 
+ i \gamma(\psi - \psi^*)
= 0.  
\label{NLS2}
\end{equation}
This 
is the form of the parametrically driven, damped NLS that 
we will be working with 
in this paper. 
The stable background solutions of Eq.(\ref{NLS2}) are simply
$\psi _{\mbox{flat}}= \pm 1$.

Solutions to Eq.(\ref{NLS2}) with  $|\psi| \to 1$\ as $|x|\to \infty$\ 
can either be topological $\left[\psi(-\infty) = -\psi(\infty)\right]$\ or 
nontopological
$\left[\psi(-\infty) = \psi(\infty)\right]$. 
Eq.(\ref{NLS2}) has two explicit topological solitons. 
The first is the  N\'eel, or Ising, wall \cite{XY,Raj,Niez,Niez2,Elphick_Meron}:
\begin{equation}
\label{Neel}
\psi_N  (x) = \tanh(x),
\end{equation}
named after the N\'eel wall in magnetism, which is a domain wall 
with the magnitude of the magnetisation vector vanishing   at its centre.
The second topological solution, which exists only for $\gamma = 0$,  
is usually referred to as the
Bloch wall \cite{Sarker,Montonen,Niez,Niez2}:
\begin{equation}
\label{Bloch}
\psi_B (x) = \tanh(Bx) \pm i \kappa_0 \,   \mbox{sech}(Bx).
\end{equation} 
(In magnetism, a Bloch wall is a domain wall  
connecting the two domains smoothly,
with the magnetisation vector  remaining nonzero everywhere.)
In Eq.(\ref{Bloch}), 
\[
B = \frac{2\sqrt{A^2-1}}{A}=\sqrt{\frac{4h}{1+h}} 
\] 
and
\begin{equation}
\kappa_0 = \frac{\sqrt{4-3A^2}}{A}= \sqrt{\frac{1-3h}{1+h}}. 
 \label{kappa_0}
\end{equation}
 The solutions obtained by multiplying $\psi_N$\ 
and $\psi_B$\ by $(-1)$\ will be
called antiwalls, or antikinks. 

The Bloch wall (\ref{Bloch})  exists in two chiralities.  
In the Bloch wall with positive imaginary part, the phase of the complex field
$\psi (x)$\ decreases (that is, the point on the unit circle moves clockwise) 
as  $x$\ varies from $-\infty$\ to $+\infty$. 
By analogy with the right-hand rule of circular motion, we refer to
this wall as the right-handed Bloch wall. 
The phase of the Bloch wall with negative imaginary part increases as $x$\ 
increases (that is, the
phase vector rotates counter-clockwise). This
corresponds to a left-handed sense of rotation, so that this wall 
will be called the left-handed Bloch wall.
The antiwall obtained by multiplying $\psi_B$
by $-1$ has obviously the same chirality as its parent wall, $\psi_B$.
 Regardless of
their chirality, Bloch walls only exist for $h < \frac 13$.

In Ref.\cite{OurPaper}, it was proved that the N\'eel wall is stable 
for all $h > \gamma$.
The stability of the Bloch wall in its entire domain of existence was 
demonstrated numerically \cite{OurPaper}.

Examples of {\it nontopological\/} dark solitons are given by stationary
complexes, or bound states, of domain walls. These can be 
formed by two dissipative N\'eel walls, or by a Bloch
and  N\'eel wall --- in the undamped situation. (See
\cite{OurPaper,OurPaper2}.)
In condensed matter physics, these nontopological solitons
describe bubbles of one thermodynamic phase in another one
\cite{bubbles}. Accordingly, we will occassionally
be referring to bound states of domain walls as solitonic bubbles.

\section{Interactions between the walls: the method}
\label{Variational}

In this paper, we will be paying special attention to the 
simplest situation where the interacting walls are initially at rest.
Choosing the origin of the coordinate axis midway 
between two Bloch or two N\'eel walls, one can verify that
the subsequent evolution will preserve this symmetric
arrangement
(see below).  
A symmetric pair of well separated walls
can be approximated by a product function  
\begin{subequations}
\label{Ansatz} 
\begin{equation}
\psi = \varphi_1(x,t) \varphi_2(x,t), 
\label{Convenience3} 
\end{equation}
where $\varphi_1$ and $\varphi_2$ represent the 
individual walls in the pair: 
\begin{eqnarray}
  \varphi_1(x,t)= \tanh{[{\cal B} (x+z)]} 
  - i\kappa_1 \mbox{sech} [{\cal B} (x+z)],
  \label{Convenience1} \\
  \varphi_2(x,t)=\tanh{[{\cal B} (x-z)]} 
+ i\kappa_2 \mbox{sech} [{\cal B} (x-z)]. 
  \label{Convenience2} 
\end{eqnarray}
\end{subequations}
In Eq.(\ref{Ansatz}),  
$\kappa_1$ and  $\kappa_2$  are time-dependent parameters accounting for 
the deformation of the walls due to their interaction.
The variable $z = z(t)$ gives half the distance between the walls;
without loss of generality, we take
$z$ to be positive.
Finally, the constant ${\cal B}$ characterises 
the width of the walls (${\cal B} = B$ for Bloch
walls and ${\cal B} = 1$ for N\'eel walls). 

Our analysis 
is based on the variational method. For this method,
we substitute the Ansatz (\ref{Ansatz}) 
 into the action integral that gives rise to Eq.(\ref{NLS2}):
\begin{equation}
\label{S_cal}
{\cal S}= \int\mathcal{L} e^{2\gamma t}\ dt, 
\end{equation}
where
\begin{widetext} 
\begin{equation}
\mathcal{L} = \mbox{Re} \int \left\{ i\psi_t \psi^* - \frac {A^2}{2}
\abs{\psi_x}^2  - \frac {A^2}{2} \abs{\psi}^4 + \abs{\psi}^2 
+ \frac{A^2-1}{2} \left[\psi^2 + 
(\psi^*)^2\right] +\frac{i\gamma}{2}
\left[\psi^2 - (\psi^*)^2\right]  - \frac{A^2}{2} \right\} dx.
\label{Lagrange}
\end{equation} 
\end{widetext}
In what follows 
we introduce a small parameter $\epsilon=e^{-2 {\cal B}
z}$.
Integrating off the explicit $x$-dependence in (\ref{Lagrange}),
produces a finite-dimensional lagrangian 
\begin{subequations} 
\label{LLL}
\begin{equation}
\label{Lag}
{\cal L}= T-V,
\end{equation}
 where 
\begin{eqnarray}
T = -\pi\dot{z} (\kappa_1 + \kappa_2)\left[1- 2(1-\kappa_1\kappa_2)(\epsilon - 2\epsilon^2)\right] 
\nonumber \\ 
- \frac{2\pi}{\cal B}
(\epsilon - 4\epsilon^2)\left[\dot{\kappa_1}(1-\kappa_2^2) +\dot{\kappa_2}(1-\kappa_1^2) \right],
\label{TT}  
\end{eqnarray}
and
\begin{widetext}
\begin{multline}
V = \frac{A^2 {\cal B}^2+ 8A^2 -12}{3 {\cal B}}(\kappa_1^2 + \kappa_2^2)  
+ \frac{2A^2}{3 {\cal B}}(\kappa_1^4 + \kappa_2^4) 
+8  \left[  (4A^2-A^2 {\cal B}^2 -4)z 
-\frac {4}{{\cal B}}(A^2-1)\right]\epsilon\kappa_1 \kappa_2    
\\
+\frac{8}{3{\cal B} } \epsilon^2 \left[ 2A^2(4{\cal B} ^2-7) 
+ 
(30A^2-5A^2 {\cal B}^2 -12)(\kappa_1^2 + \kappa_2^2) \right. \\
\left.
+ 2A^2 ({\cal B}^2-11) \kappa_1^2\kappa_2^2 
 -4A^2(\kappa_1^4+\kappa_2^4) 
+ 4A^2(\kappa_1^2+\kappa_2^2)\kappa_1^2\kappa_2^2 \right] 
\\  
+ 16\epsilon^2 z \left[ 2A^2(1-{\cal B}^2) + 
(A^2 {\cal B}^2 - 6A^2 + 4) (\kappa_1^2 + \kappa_2^2)
+ 2A^2\kappa_1^2\kappa_2^2 \right] 
+ 
\frac{4\pi\gamma}{{\cal B}} \epsilon (1-4 \epsilon) 
(\kappa_1 + \kappa_2)
(1+\kappa_1\kappa_2). 
\label{VV} 
\end{multline}
\end{widetext}
\end{subequations}
[In (\ref{TT}), the overdot indicates differentiation with respect to
$t$.]
The
stationary action principle $\delta {\cal S}=0$ yields then 
the Euler-Lagrange equations
for $\kappa_{1,2}$ and $z$. In deriving (\ref{TT})-(\ref{VV}) 
we neglected all powers of $\epsilon$ higher than $\epsilon^2$.
One can readily check that such terms would produce
only higher-order corrections in the resulting equations of motion.

Some comments must be made regarding the Ansatz (\ref{Ansatz}).
After a variational Ansatz has been substituted in the 
corresponding field Lagrangian and
the $x$-dependence integrated away, all dynamical variables
(or possibly their combinations) can
be grouped into canonically-conjugate pairs.
(See \cite{MalomedReview} for review and references.)
In particular, if the waveform consists of {\it two\/} solitary waves
and the separation between two constituents, $2 z$, is 
chosen as one of the variables,  the conjugate momentum 
  is the phase gradient. This fact is well known in the case 
  where the constituents are bell-shaped (``bright") solitons 
  (see e.g. \cite{YanMei,Variational}).
That it remains true in a more
general situation can be seen from an  analogy with
quantum mechanics where the momentum of the system is given by an eigenvalue
of the operator $\frac{1}{i} \partial_x$.
(Since the eigenvalue has to be real, the operator acts just on the phase
of the eigenfunction while the modulus is taken to be constant.) 
Another simple
analogy arises if we write $\psi=\sqrt{\rho} e^{i \chi}$; this 
polar decomposition casts Eq.(\ref{NLS}) with $h=\gamma=0$ in the
form of equations of gas dynamics,
where $\rho$ is the density of the gas and $\partial_x \chi$
is its velocity,  proportional to the momentum.

In the case of bright solitons,
 each constituent soliton in the
Ansatz is usually multiplied by $e^{ik(x \pm z)}$
(see e.g. \cite{YanMei,Variational}); then,  the momentum conjugate to $z$
is $k$.
However, in the case of kinks this simple
recipe
would  violate the boundary conditions at infinity. For this reason we have
introduced
the phase gradient by allowing the imaginary parts
$\kappa_{1,2} \, \mbox{sech}\, [{\cal B} (x \pm z)] $\ 
of the kinks to be variable. 
The imaginary parts decay as
$|x| \to \infty$ and therefore making them
variable is compatible with the boundary conditions.
Below, we will show that $\kappa_1$ and $\kappa_2$ 
are indeed momenta canonically conjugate to $z$.

Another reason for introducing the phase gradient in this
way is a simple  interpretation of the finite-dimensional
momenta  $\kappa_1$ and $\kappa_2$
in terms of the original fields $\psi(x,t)$.
 Indeed, these
variables coincide, modulo a numerical coefficient,
with the {\it field momenta\/} of the two walls. 
The field momentum integral has the form
\begin{equation}
\label{momentum}
P = \frac i2 \int \left( \psi_x^* \psi - \psi_x \psi^*\, \right) dx.
\end{equation}
For the undamped NLS [equation (\ref{NLS2}) with $\gamma=0$], 
the integral (\ref{momentum})
is conserved. The field momentum of the N\'eel wall (\ref{Neel}) equals zero, 
while 
that of the Bloch wall
(\ref{Bloch}) is given by $P_{\rm B} = \mp \pi \kappa_0$. [The
same sign convention is used here as for Eq.(\ref{Bloch}) --- the 
top sign applies for the right-handed wall, 
while the bottom sign corresponds to the left-handed wall.]
It is not difficult to check that the  field momenta of
 the perturbed walls $\varphi_1$ and $\varphi_2$
 in the configuration (\ref{Ansatz}) are
  $P[\varphi_1]=\pi \kappa_1$
 and $P[\varphi_2]=-\pi \kappa_2$.

In symmetric situations that we are concerned with, the 
variables 
 $\kappa_1$\ and $\kappa_2$ (both of which are conjugate to the 
same coordinate
variable $z$) will eventually turn out to 
be related due to the equations of motion. 
The calculations simplify, however, if  
this relationship is established yet at the level of
the lagrangian (\ref{LLL}), that is, before
the equations of motion have been derived.

The relationship between $\kappa_1$ and $\kappa_2$ 
 depends on the type of walls being considered. 
If the initial configuration consists of two well-separated N\'eel walls
at rest, 
or two quiescent distant Bloch walls of {\it opposite\/}  chirality,
we have a symmetry $\psi(x,0)=\psi(-x,0)$. Since 
equation (\ref{NLS2}) is parity-invariant, the subsequent evolution
will preserve this symmetry.
Substituting the Ansatz (\ref{Ansatz}) into 
 $\psi(x,t)=\psi(-x,t)$, we get $\kappa_1=\kappa_2$.
The initial value of $\kappa(t) $\ should be chosen 
close to zero for N\'eel walls and close to $\pm \kappa_0$
 for Bloch walls [with the sign of $\kappa(0)$ depending 
on whether the wall placed on the right is right- or left-handed.]

If the initial configuration consists of
two quiescent  Bloch walls of {\it like\/}  chirality,
the relationship between $\kappa_1$ and $\kappa_2$ is
not so trivial to establish. The reason is that equation (\ref{NLS2})
{\it does  not\/} preserve the symmetry $\psi^*(x,0)=\psi(-x,0)$ of the 
initial condition. In this case we have to
appeal to the 
field-momentum considerations rather than symmetries.
Substituting  the Ansatz (\ref{Ansatz}) into the integral
(\ref{momentum}), we find that $P[\psi] = \pi (\kappa_1 - \kappa_2) [1 +
\mathcal{O}(e^{-2 {\cal B} z})]$. 
On the other hand, the total momentum of two like-chirality 
walls which are initially at rest and far away from each other,
should be near $ \mp 2 \pi \kappa_0$. 
Therefore, for moving walls the 
quantity $\kappa_1-\kappa_2$ must not be different 
from $ \mp 2\kappa_0$ by more than ${\cal O}(e^{-2Bz})$. 
This is accomplished by letting $\kappa_1 = \mp \kappa_0 + q(t)$\ and
$\kappa_2 = \pm \kappa_0 + q(t)$. 
When $z$ is large, the perturbation $q(t)$ is small ---
but not necessarily as small as $\epsilon$.

The Euler-Lagrange equations corresponding to
Eqs.(\ref{S_cal}),(\ref{LLL}) are 
\begin{subequations}
\label{EL_0}
\begin{equation}
\frac{\partial T}{\partial \kappa} - \frac{\partial V}{\partial \kappa} -
\frac{d}{dt} \frac{\partial T}{\partial \dot{\kappa}} - 2\gamma
\frac{\partial T}{\partial \dot{\kappa}} = 0,
\label{First EL}
\end{equation}
\noindent and
\begin{equation}
\frac{\partial T}{\partial z} - \frac{\partial V}{\partial z} -
\frac{d}{dt} \frac{\partial T}{\partial \dot{z}} - 2\gamma
\frac{\partial T}{\partial \dot{z}} = 0.
\label{Second EL}
\end{equation}
\end{subequations}
In Eq.(\ref{First EL}), we assumed the situation of 
$\kappa_1 = \kappa_2 \equiv \kappa$
(that is, two N\'eel or two 
oppositely-handed Bloch walls). In the other case, i.e. when we have two
 Bloch walls of like chirality, $\kappa$\ should be  replaced by $q$\ in Eq.(\ref{First EL}).
The integrals $T$\ and $V$\ can both be found explicitly and simplified  
assuming wide separation of the walls.

Finally, we note that  the variational method 
is not well suited for the analysis of the
Bloch-N\'eel interaction. The reason for this is that 
the Bloch and N\'eel walls have different widths, 
leading to terms in the Lagrangian
which are not periodic along the imaginary axis on the plane of complex $x$;
as a result, the Lagrangian
cannot be obtained in closed form. (When the walls have the same width, 
the integrals can all 
be evaluated by integration along the
rectangular contour with one side on the real axis and the other on the
line $\mbox{Im} \, {x} = \pi /\cal B$.)
Consequently, we had to resort to a different approach 
for the analysis of the 
N\'eel-Bloch interaction (see paper II of this project \cite{BW2}.)
The resulting phenomenology is also very different \cite{BW2}.

\section{Two N\'eel walls}
\label{NeelVar}

In this case, we let ${\cal B}=1$ and $\kappa_1 = \kappa_2 \equiv
\kappa$.  
In equations (\ref{EL_0})
we discard products of powers of small quantities $\epsilon$ and $\kappa$ 
for which 
there are larger counterparts. For instance, we drop a term proportional
to  $\epsilon^2 \kappa$
from an equation which 
already has a term $\epsilon \kappa$ with a nonsmall coefficient.
Here we keep in mind that in some cases the damping $\gamma$ and the 
difference $|A^2-\frac43|$ can be small parameters, hence 
we cannot drop $\epsilon^2 \kappa$ in favour of the term $\gamma \epsilon \kappa$
or $(A^2-\frac43) \epsilon \kappa$. 
With these simplifications, equations (\ref{First EL})-(\ref{Second EL})
take the form
\begin{subequations} 
\label{general}
\begin{eqnarray}
\pi\dot{\epsilon}  
= 12 \left(A^2 - \frac 43 \right) \kappa \epsilon 
 \nonumber \\ + 32 \pi \gamma \epsilon^2 \kappa^2 +
\frac{16}{3} A^2 \kappa^3 \epsilon 
 - 64(A^2-1) \kappa \epsilon^2, 
\label{31} \\
\pi(\dot{\kappa} + 2\gamma\kappa)  
= 32 A^2 \epsilon ^2  \nonumber \\ 
- 24 \left(A^2 - \frac 43\right) \kappa^2\epsilon z
+ 32 (A^2-1) \epsilon \kappa^2. 
\label{30} 
\end{eqnarray}
\end{subequations}
The subsequent analysis depends on the relation 
between the damping and driving in the system.

\subsection{                           
$\gamma={\cal O}(1)$}
\label{both_large}

First we consider the case where $\gamma$ is not small.
Assume, in addition, that
$A^2$\ is not close to $\frac 43$. (Later
in this subsection we will
explore the situation where  $|A^2-\frac43|$ is small.)
In this case 
the terms in the second lines of (\ref{31}) and (\ref{30}) are
smaller than those in the first lines and can be neglected. 
The 
equations  simplify to
\begin{subequations} 
\label{gA_large}
\begin{eqnarray}
\pi \dot{\epsilon}  
= 12 \left(A^2 - \frac 43\right) \kappa \epsilon,
\label{gA_large_1} \\
\pi(\dot{\kappa} + 2\gamma\kappa)  = 32 A^2 \epsilon ^2.
\label{gA_large_2} 
\end{eqnarray}
\end{subequations}
Since  $\kappa$ is small, the variable 
$\epsilon(t)$\ varies very slowly, 
$\dot{\epsilon}/\epsilon \sim \kappa$. 
On the other hand, the variable $\kappa(t)$ will initially
change on a much faster scale.
 Within the time $ \Delta t \sim \gamma^{-1}$\ it will ``zap" onto 
 the nullcline
\begin{equation}
2 \pi \gamma\kappa = 32 A^2 \epsilon^2, 
\label{parabola}
\end{equation}
 after which the point 
$(\epsilon,\kappa)$ will be 
slowly moving along this parabola. 
According to Eq.(\ref{gA_large_1}), 
it will move towards greater $\epsilon$ (i.e. the 
separation $2z$ will decrease)
if 
$A^2> \frac43$ and to smaller $\epsilon$ if  $A^2 < \frac43$. That is,
the walls will attract if  $A^2> \frac43$ and repel if $A^2 < \frac43$.

Note that this criterion
is consistent with results for the undamped, undriven case 
($\gamma=0$, $A^2=1$) 
where the
dark solitons are known to repel \cite{ZakShab,KivsharReview}. 
 
\begin{figure}

\includegraphics[height = 2.0in, width = 0.5\linewidth]{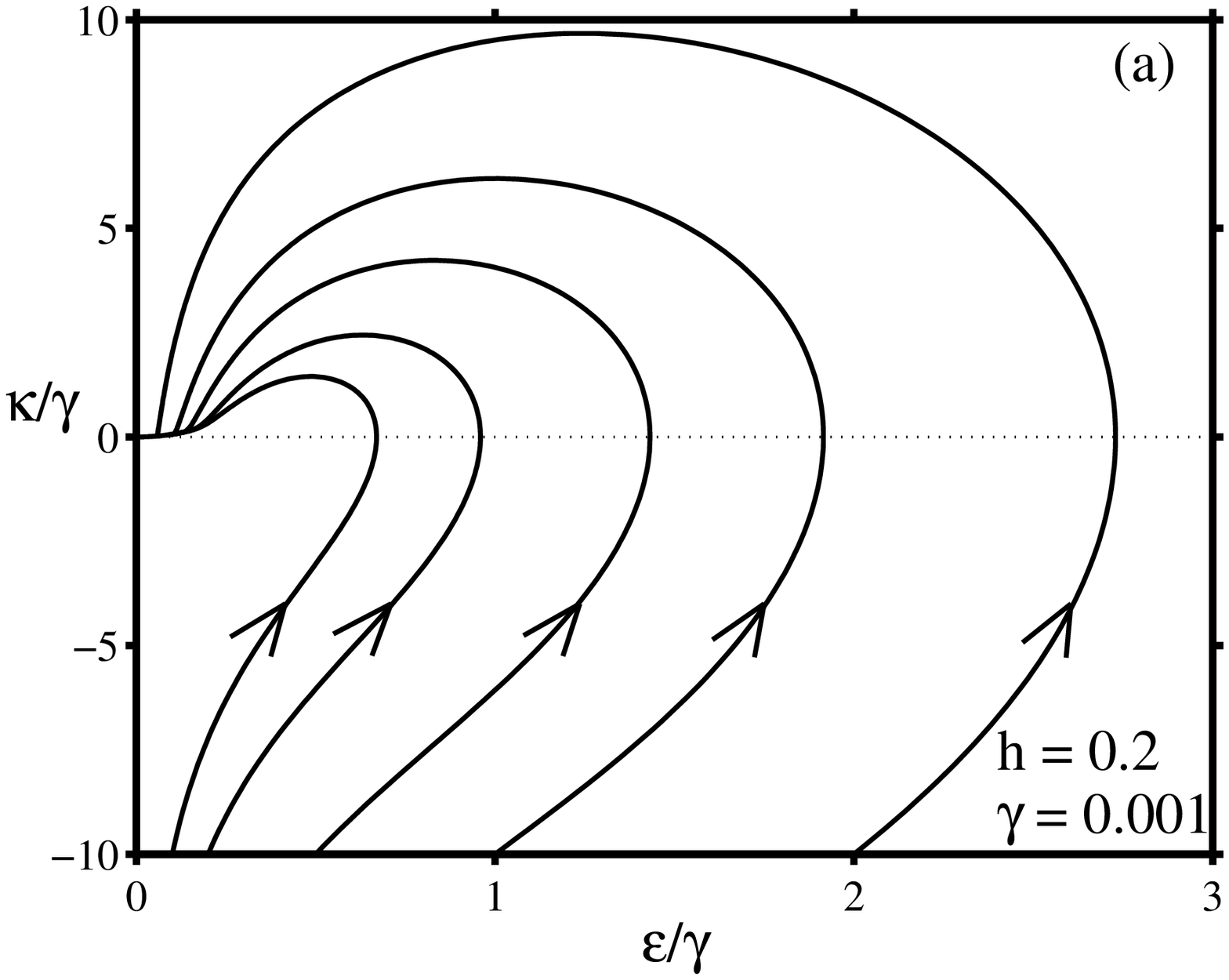}
\includegraphics[height = 2.0in, width = 0.5\linewidth]{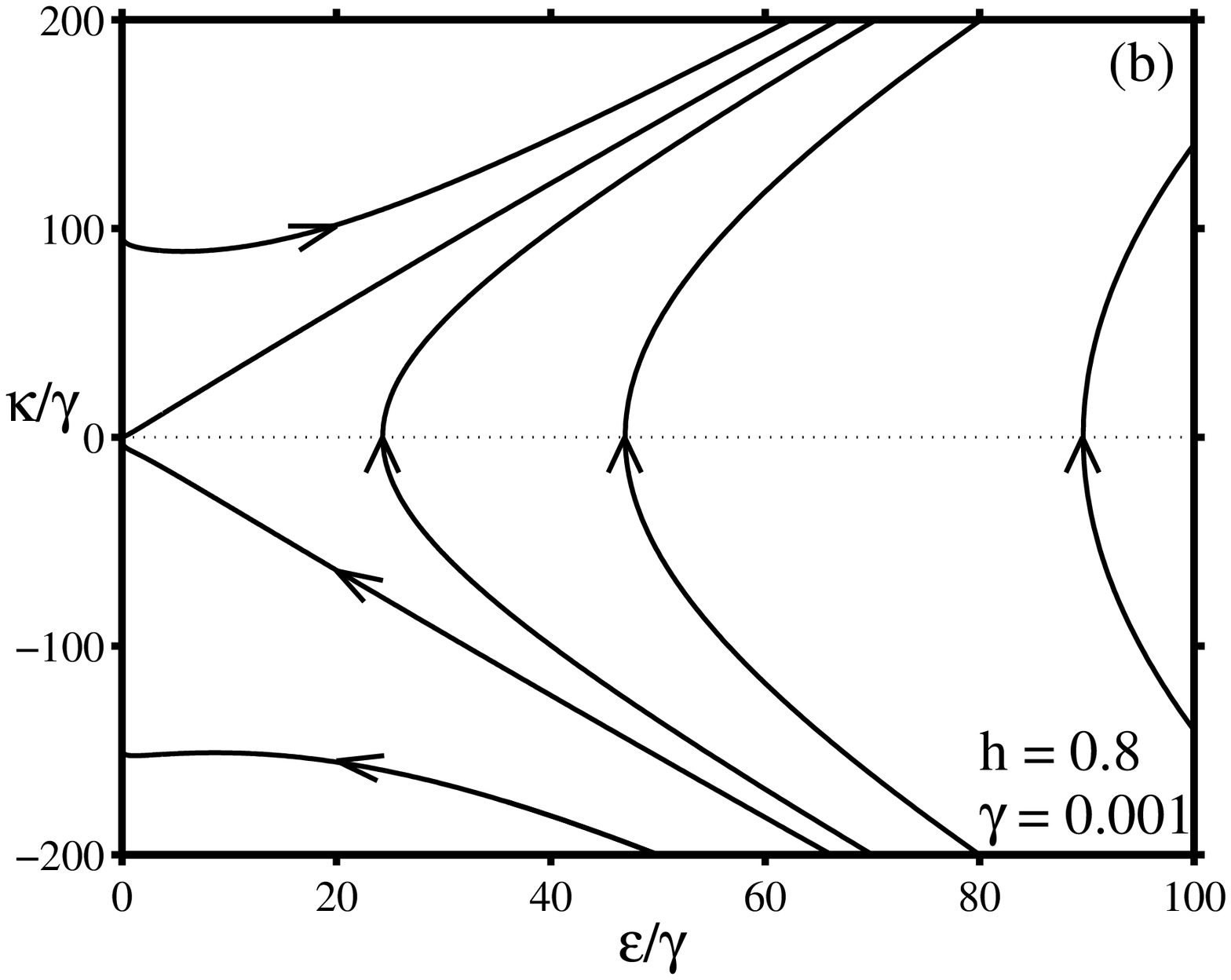}
\includegraphics[height = 2.0in, width = 0.5\linewidth]{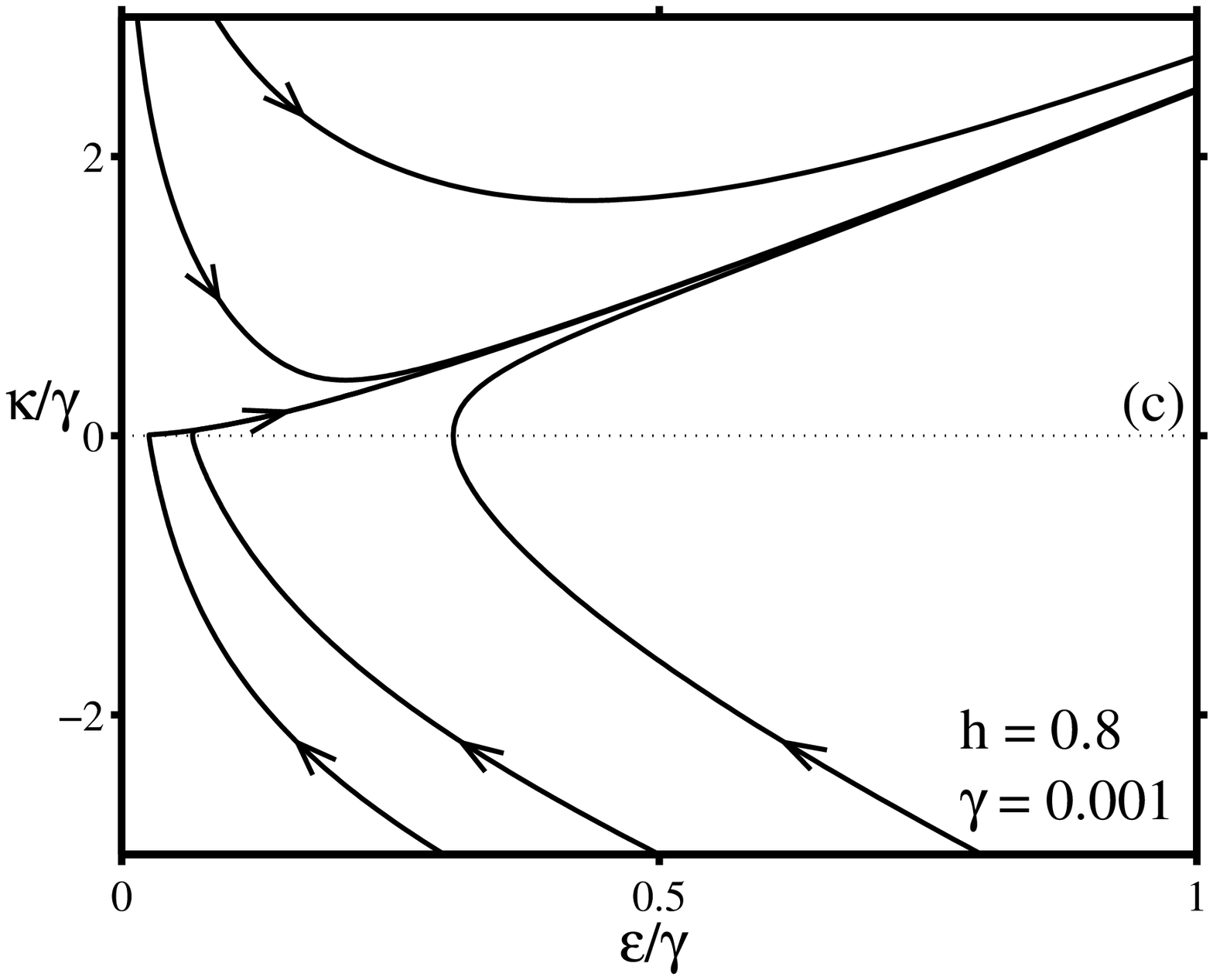}

\caption{\sf  Vector field (\ref{small_gamma}). 
(a) $A^2 < \frac 43$\ (here $A^2 = 1.2$); (b) $A^2 > \frac 43$\ 
(here $A^2 = 1.8$). (c) gives a blow-up of a small neighbourhood of the
origin in (b).}
\label{IBfig1}
\end{figure}

Since the above criterion has been derived under the assumption that
 $A^2$ is not very close to $\frac43$, the value
 $A^2=\frac43$ provides only a rough watershed between the two 
 types of interaction.
 In order to establish a more accurate borderline, 
 we zoom in on a narrow strip along the
 line $A^2=\frac43$
 on the $(\gamma, A^2)$-plane. Assuming that 
   $|A^2-\frac 43|$ is small (while $\gamma$ is not),
 Eq.(\ref{gA_large_1}) should be replaced by
\begin{equation}
\pi \dot{\epsilon} =  
12 \left(A^2 - \frac 43\right) \kappa \epsilon 
+  \frac{64}{9}  \kappa^3 \epsilon 
 - \frac{64}{3} \kappa \epsilon^2. 
\label{310} 
\end{equation} 
Substituting (\ref{parabola}) into (\ref{310}) and dropping an
${\cal O}(\epsilon^7)$ term, we obtain 
a one-dimensional dynamical system
\begin{equation}
\pi {\dot \epsilon} =
\frac{ 256 }{\pi \gamma} 
\left( A^2-\frac43 -\frac{16}{9} \epsilon \right) \epsilon^3.
\label{BSg}
\end{equation}
When $A^2< \frac43$, $\epsilon(t)$ tends to zero for all $\epsilon(0)$. 
For $A^2 > \frac43$, 
the 
system  (\ref{BSg})  has a stable fixed point at 
\begin{equation}
\label{separa}
\epsilon = \frac{9}{16} \left( A^2-\frac43 \right).
\end{equation}
This fixed point corresponds to a stable bound state of two N\'eel
walls.

When $A^2$ approaches  $\frac43$, the distance $2z=- \ln \epsilon$
between the walls in this bound state tends to infinity
and so the bound state does not exist for $A^2$ smaller than $\frac43$.
This means that 
$A^2=\frac43 $ is, in fact, an {\it accurate\/} borderline between the 
two types of behaviour.
For $A^2< \frac43$, two  walls repel whereas for  
$A^2> \frac43$, the  walls attract  --- except for
 $A^2$ close to
$\frac43$, in which case they form a stable bound state.

We should emphasise here that our present conclusions pertain 
only to distant walls.
In order to extend our understanding of the N\'eel
wall dynamics beyond the limit of  
{\it very\/} small $\epsilon$, one has to analyse the  
 dynamical system (\ref{general}) without neglecting any 
 powers of $\epsilon$ and $\kappa$
  in it. This will be done numerically in 
 section \ref{numerical_FP} below.
We will show, in particular, that in addition to the stable
bound state of two walls, there is also an unstable complex ---
at a shorter distance.

\subsection{ $|A^2-\frac 43|={\cal O}(1)$; small $\gamma$ }

\begin{figure*}
\includegraphics[height = 2.0in, width = 0.45\linewidth]{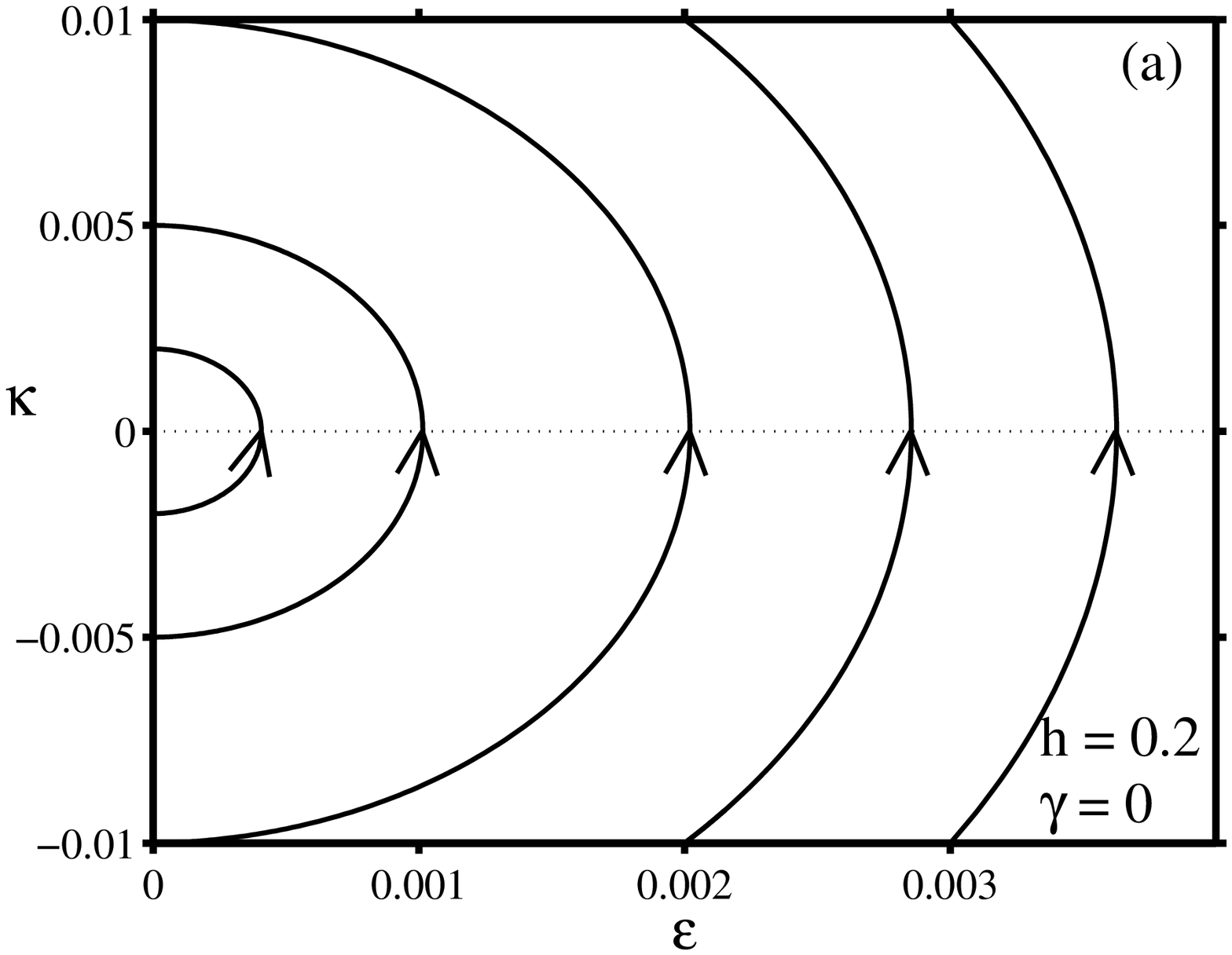}
\includegraphics[height = 2.0in, width = 0.45\linewidth]{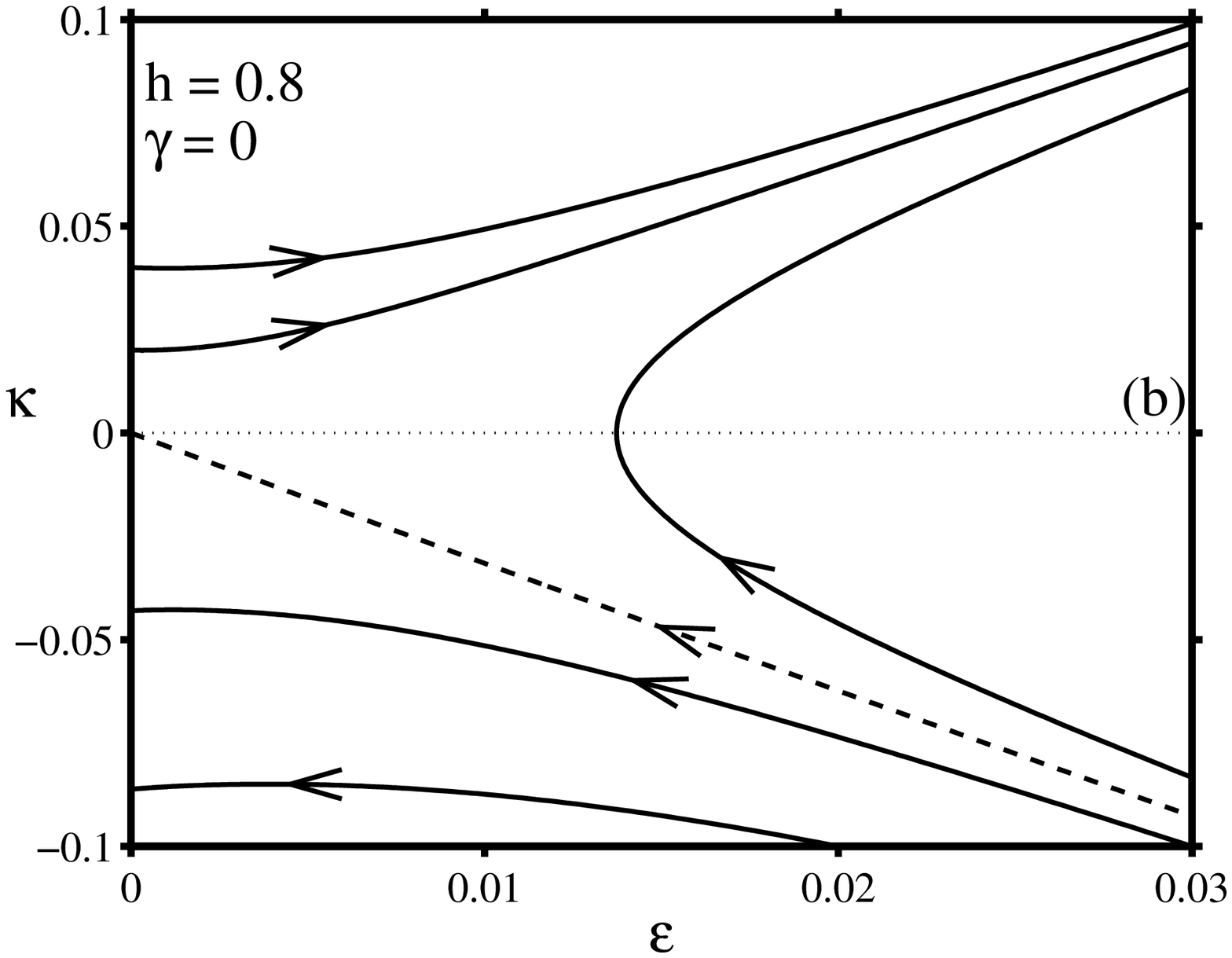}
\caption{\sf  Vector field (\ref{small_gamma}) with $\gamma=0$. 
 (a) $A^2 < \frac 43$;
(b) $A^2 > \frac 43$.}
\label{IBfig11}
\end{figure*}

When $\gamma$ is small, the system (\ref{general}) becomes
\begin{subequations}
\label{small_gamma}
\begin{eqnarray}
\pi {\dot \epsilon}= 12 \left( A^2-\frac43 \right) 
\kappa \epsilon,
\label{small_gamma_1} \\
\pi ({\dot \kappa} + 2 \gamma \kappa) = 32 A^2 \epsilon^2 
-24 \left(A^2-\frac43\right) \kappa^2 \epsilon z.
\label{small_gamma_2}
\end{eqnarray}
\end{subequations}
For either 
sign of $A^2-\frac43$, the origin is the only fixed point in the system.
Noting that the $\kappa$-axis ($\epsilon=0$) is an invariant manifold 
flowing into the
fixed point and calculating the index of the vector field (\ref{small_gamma}) 
along the semicircle 
$\epsilon^2+\kappa^2=R^2$, $\epsilon \geq 0$,
with $R<1$,  we conclude that 
the fixed point is a stable node
for $A^2< \frac43$ and a saddle for $A^2> \frac43$.
Hence, for $A^2< \frac43$, all trajectories tend to
the origin as $t \to \infty$ [Fig.\ref{IBfig1}(a)], i.e., the walls repel.

For $A^2> \frac43$,
all trajectories fly
away to infinity [Fig.\ref{IBfig1}(b)] implying the attraction of the walls.
One may wonder whether some trajectories 
coming from the lower half of the phase plane, could flow into
the origin (tangent to the vertical axis.) 
A simple analysis 
of trajectories
in a small neighbourhood of the origin 
shows that such behaviour is, in fact, 
impossible. Trajectories which seem to be approaching  the 
point $\epsilon=\kappa=0$ close to the $\kappa$-axis, do  turn away 
when they are a small distance away from the origin.
This is illustrated by
Fig.\ref{IBfig1}(c) which gives a blow-up of
 a small neighbourhood of the fixed point.

In the next  subsection,  we will zoom in on a
small neighbourhood of $A^2 = \frac43$ to draw a sharper boundary 
between domains of the opposite types of interaction
--- similar to the way we have done it for large $\gamma$.
We should also mention here that, as in the previous
subsection, our current conclusions pertain to walls
with large separations.

The undamped case, $\gamma =0$, is  
exceptional; in this case, the entire $\kappa$-axis is a line of fixed 
points --- they represent pairs of infinitely
separated walls moving at constant velocities.
For $A^2< \frac43$, the points with
  $\kappa>0$ are stable and those with  $\kappa<0$ unstable. 
Trajectories have the form of arcs 
starting  on the lower semiaxis and flowing into points on 
the upper semiaxis: $\epsilon(t) \to 0$,
 $\kappa(t) \rightarrow \kappa_* > 0$\  
 as $t \rightarrow \infty$ [Fig.\ref{IBfig11}(a)]. 
 The 
separation between  the walls 
corresponding to the points on the upper semiaxis
grows linearly: 
$z \rightarrow - \frac {6}{\pi} \left( A^2 - \frac43 \right)
\kappa_* t$; that is, 
the evolution produces
two N\'eel walls moving away from each other at a constant speed.

In the undamped case with
  $A^2> \frac43$, the stable points are those with
  $\kappa <0$ and unstable those with  $\kappa>0$. 
  Since ${\dot \kappa}>0$ on the horizontal axis (where
  $\kappa=0$), trajectories 
starting on the upper semiaxis cannot cross 
to the negative-$\kappa$ half-plane and have to escape to the 
positive-$\kappa$ infinity:
 $\kappa \to \infty$,
$\epsilon \to \infty$ as $t \to +\infty$.
The same is true for trajectories that have
arrived from the negative-$\kappa$ half-plane.
Invoking  the reversibility of the system (\ref{small_gamma})
with $\gamma=0$, i.e. making use of the invariance under $t \to -t$, $\kappa \to
-\kappa$,
completes the phase portrait [Fig.\ref{IBfig11}(b)].

To interpret the portrait, we note that
$\kappa$ is proportional to ${\dot z}$, the velocity 
of the wall
[see Eq.(\ref{small_gamma_1})].
According to Fig.\ref{IBfig11}(b), two walls with zero initial velocities 
[$\kappa(0)=0$] will attract
[$\epsilon(t)\rightarrow \infty$\ as $t \rightarrow \infty$]. 
The same applies, naturally,  to the walls 
whose initial velocities are directed towards
each other [$\kappa(0) > 0$], and to walls with small outward 
initial velocities
[$\kappa(0) < 0$].    
However, walls with sufficiently
large outward initial velocities  will diverge 
[$\epsilon(t)\rightarrow 0$], with the separation growing
linearly: 
$z(t) \rightarrow -\frac {6}{\pi}\left(A^2 - \frac 43 \right)\kappa_* t$\ 
as $t \rightarrow \infty$. 
(Here $\kappa_* < 0$.) The dashed line in Fig.\ref{IBfig11}(b) is a 
separatrix between initial conditions giving rise to the 
above two scenarios.

Finally, it is worth mentioning that when $\gamma=0$,
equations (\ref{EL_0}) with $\kappa_1=\kappa_2=\kappa$
and ${\cal B}=1$ form a hamiltonian system 
(with a nonstandard bracket):
\begin{equation}
 \dot{z}= \frac{1}{\pi \tau} \frac{\partial H}{\partial \kappa},
\quad
 \dot{\kappa}= -\frac{1}{\pi \tau} \frac{\partial H}{\partial z},
\label{H_eqs_N}
\end{equation}
where 
\begin{equation}
\tau(\epsilon, \kappa)=1+2 \epsilon(1-14\epsilon +\kappa^2 +10 \epsilon
\kappa^2)>0, 
\label{tau}
\end{equation}
and the Hamiltonian 
\begin{multline}
H(\epsilon,\kappa)=
(1-3h)\kappa^2 -\frac23 A^2 \kappa^4 +4 \epsilon \kappa^2 
[(1-3h)z+4h]
\\
+ \frac83 \epsilon^2 [3A^2 +(12-25A^2)\kappa^2 +14 A^2 \kappa^4 -4A^2
\kappa^6]
\\
+16\epsilon^2 z \kappa^2 (5A^2-4-A^2 \kappa^2).
\nonumber
\end{multline}
This observation endows $\kappa$ with a simple 
interpretation of the momentum canonically conjugate 
to $z$ ---
in agreement with  qualitative arguments 
in section \ref{Variational}.

\subsection{ Small $|A^2 - \frac 43|$; small $\gamma$}
\label{small_small}

Finally, let us assume that {\it both\/} $|A^2 - \frac 43|$\ 
and $\gamma$\ are small.
Here, equations (\ref{general}) are 
to be replaced by 
\begin{subequations} 
\label{revvy}
\begin{eqnarray}
\pi\dot{\epsilon}  
= 12 \left(A^2 - \frac 43 \right) \kappa \epsilon 
  +
\frac{64}{9}  \kappa^3 \epsilon 
 - \frac{64}{3} \kappa \epsilon^2, 
\label{revi} \\
\pi (\dot{\kappa} + 2\gamma\kappa)  
= \frac{128}{3} \epsilon ^2 + \frac{32}{3}  \epsilon \kappa^2. 
\label{reviso} 
\end{eqnarray}
\end{subequations}
Note that we have neglected the term proportional to 
$z=-\frac12 \ln \epsilon$ in the second line
of (\ref{30}). When $A^2-\frac43<0$, this is 
justifiable as this term is of the same sign as
the last  term in the right-hand side of (\ref{30})
and so it cannot alter the dynamics qulitatively.
The validity of this approximation for
 $A^2-\frac43>0$ will be established below.
  
The dynamics is influenced by 
nontrivial fixed points
which arise as intersections of the nullclines
of the system (\ref{revvy}):
\begin{subequations} 
\label{cline}
\begin{eqnarray}
\kappa \epsilon \left( 12 \mu  + \frac{64}{9} \kappa^2 
 - \frac{64}{3}  \epsilon \right)=0, 
\label{cline1} \\
  \frac{128}{3} \epsilon ^2 + \frac{32}{3} \epsilon \kappa^2- 2 \pi \gamma \kappa=0.
\label{cline2}
\end{eqnarray}
\end{subequations}
Here we have introduced 
\begin{equation}
\mu=  A^2-\frac43. 
\label{muA}
\end{equation}

 The nullcline (\ref{cline2}) emanates
out of the origin as a parabola: 
$\kappa \to \frac{64}{3\pi \gamma} \epsilon^2$ as
$\epsilon \to 0$; 
turns back 
and escapes to infinity asymptotic
to the positive $\kappa$-axis: 
$\kappa \to  \frac{3 \pi \gamma}{16} \epsilon^{-1}$
as $\epsilon \to 0$.
Importantly, it has
no points in the $\epsilon > 0$, $\kappa < 0$ quadrant.
The nullcline (\ref{cline1}) consists (apart from
the vertical
and horizontal axes) of a parabola lying on a side: $\epsilon=\frac34 \mu + \frac
13 \kappa^2$. Intersections of these nullclines 
are easy to determine and visualise, and
the stability of the fixed points can be classified by 
straightfoward index arguments.

\begin{figure}

\includegraphics[height = 2.0in, width = 0.5\linewidth]{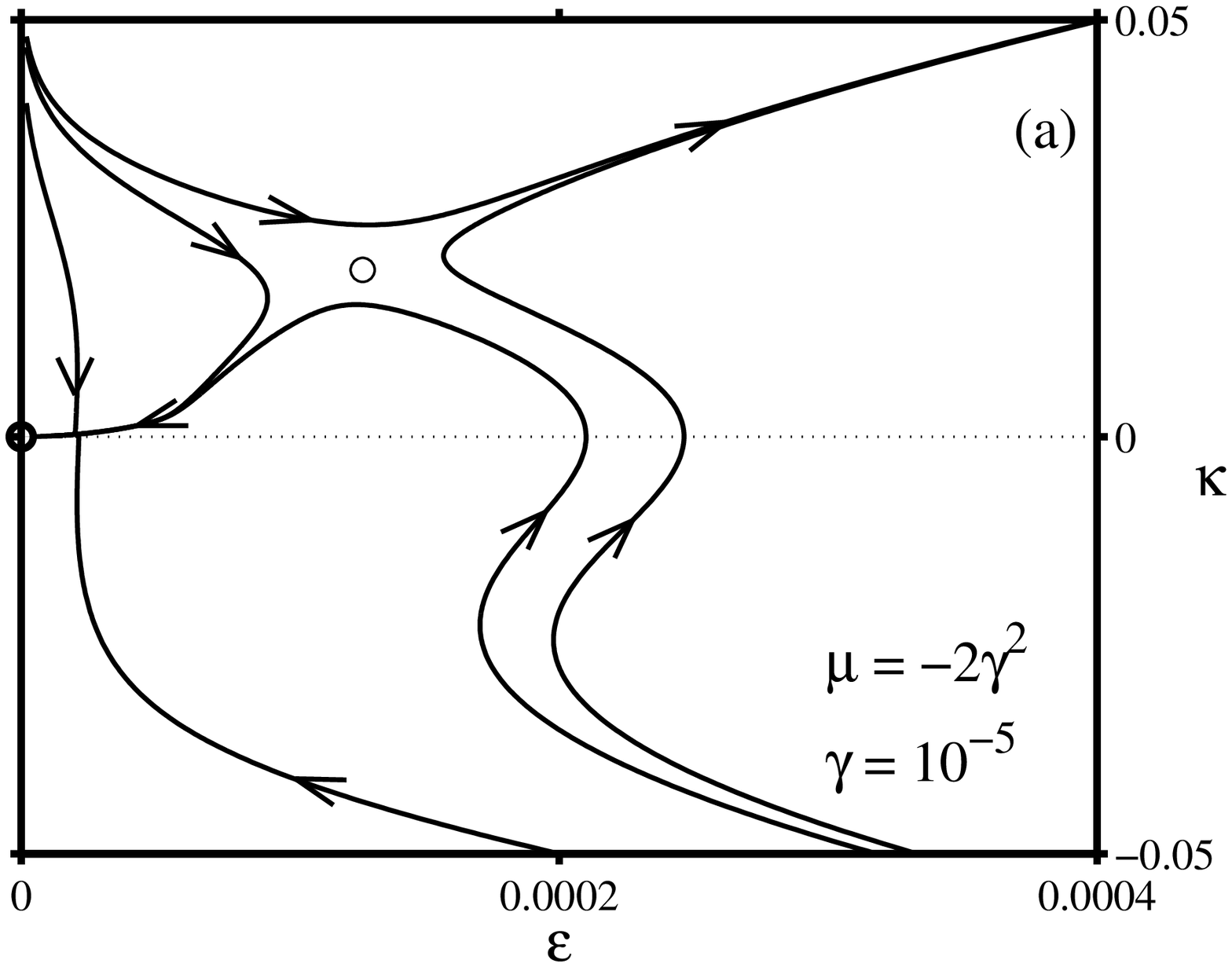}
\includegraphics[height = 2.0in, width = 0.5\linewidth]{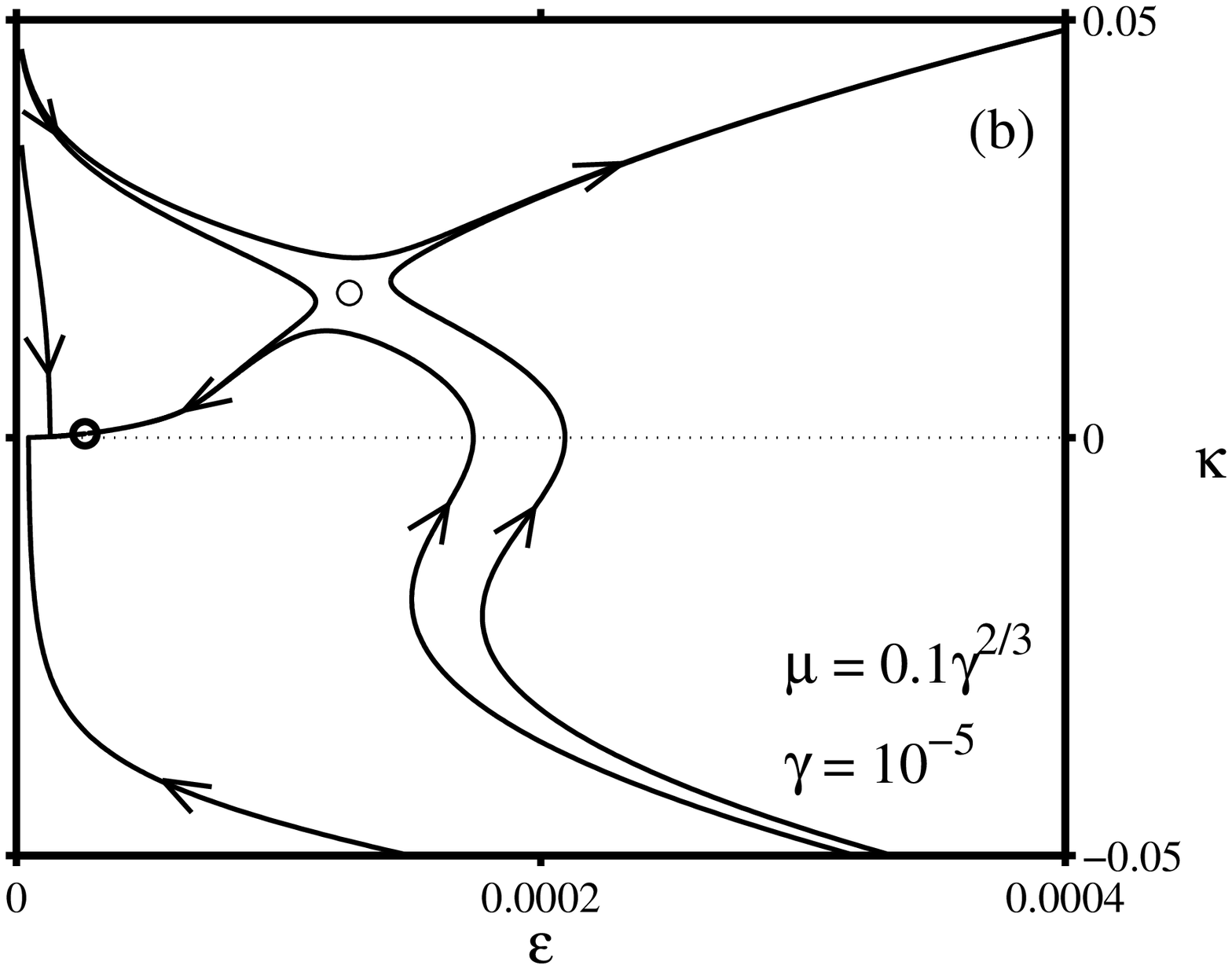}
\includegraphics[height = 2.0in, width = 0.5\linewidth]{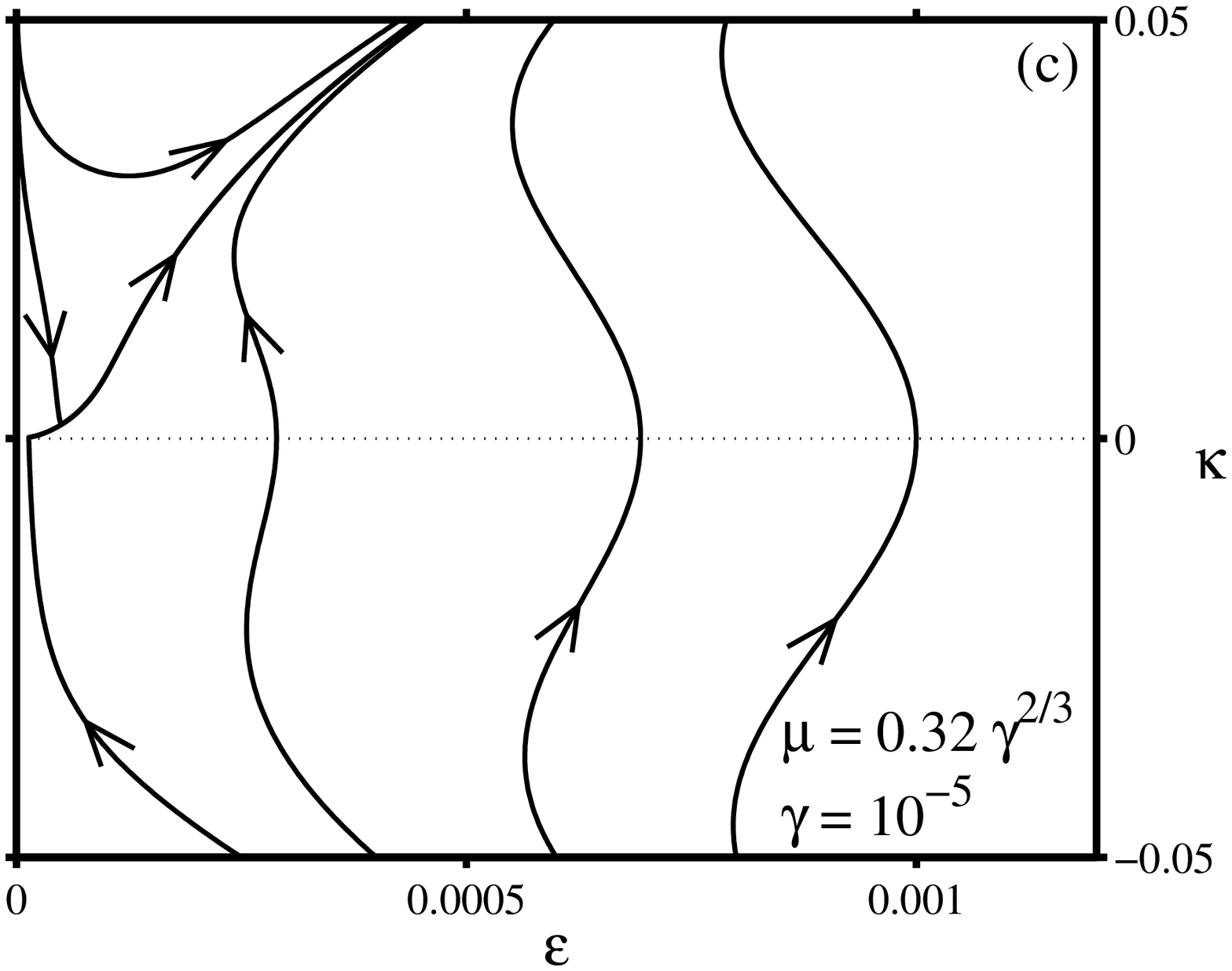}

\caption{\sf  The vector field (\ref{revvy}).
(a)  $\mu< 0$ (small initial conditions
are attracted to a stable fixed point at the origin; larger 
initial conditions give rise to trajectories flowing 
to infinity).
(b)  $0 < \mu < 0.3048 \gamma^{2/3}$
(initial conditions with small $\epsilon$ and $\kappa$ are attracted
to a stable fixed point away from the origin; larger 
initial conditions give rise to unbounded trajectories). 
(c) $\mu> 0.3048 \gamma^{2/3}$ (no stable fixed points,
all trajectories flow to infinity).
The stable fixed point is marked by a solid blob; the unstable 
one by an open circle.
}
\label{fig2}
\end{figure}

When $\mu<0$, there is only one  nullcline intersection
 in the positive-$\epsilon$
halfplane.
It is not difficult to check that this fixed point
is a saddle. The origin is also a fixed point ---
a stable node. The portrait is in Fig.\ref{fig2}(a).
Distant walls
[more precisely, walls with initial conditions
lying below the stable manifold of the
saddle in Fig.\ref{fig2}(a)] repel: $\epsilon(t) \to 0$ as $t \to
\infty$. On the other hand, nearby walls (more precisely, those 
above the stable manifold) attract: $\epsilon(t) \to \infty$.

 As $\mu$ grows through zero, another fixed point
appears from the negative-$\epsilon$ half-plane. 
An exchange of stabilities 
 occurs at $\mu=0$: the 
nontrivial point 
appearing for $\mu>0$ is a stable node 
whereas the origin becomes a saddle.
The nearby walls  attract while distant walls form 
a stable bound state [Fig.\ref{fig2}(b)].
As $\mu$ is increased further,
the two nontrivial fixed points merge in a saddle-node bifurcation. 
When $\mu>\mu_{sn}$, the origin 
persists as the only fixed point in the system
(an unstable one). All walls attract [Fig.\ref{fig2}(c)].

Note that the logarithmic term in (\ref{30}) that was
omitted in (\ref{reviso}), becomes of the same order as 
 the last  term in the right-hand side of (\ref{30})
 only if $z \mu \sim 1$, i.e. when $\epsilon$
 is as small as $e^{-2/\mu}$. 
 However, since 
 the nullcline (\ref{cline1}) is bounded from the vertical
 axis by the inequality $\epsilon \geq
 \frac{9}{16} \mu$, there can be no fixed points with
  $\epsilon \sim e^{-2/\mu}$. Therefore the logarithmic term
  can have no significant effect on the dynamics
  and its 
  omission is totally legitimate.

To find the threshold  $\mu_{sn}(\gamma)$, we eliminate $\kappa$ 
between (\ref{cline1}) and (\ref{cline2}). The resulting 
equation can be written as
\begin{equation}
    \epsilon \frac{7\epsilon- \frac{27}{16} \mu } 
  {\sqrt{3\epsilon- \frac{27}{16} \mu  }}
  =\frac{3 \pi}{16} \gamma.
\label{pi9}
\end{equation}
If $\mu$ is positive, the function $F(\epsilon)$ 
in the left-hand side of (\ref{pi9}) grows to infinity 
 as $\epsilon \to \frac{9}{16}\mu $ and $\epsilon \to \infty$, and 
 has a minimum at $\epsilon=\epsilon_{min} =\frac{1}{126}(31+\sqrt{457})
 \mu$. The minimum value is $F(\epsilon_{min}) = F_0 \mu^{3/2}$, 
 where $F_0$ is a numerical coefficient.
 Consequently, Eq.(\ref{pi9})
 has no roots if $ \mu> (3 \pi / 16 F_0)^{2/3} \gamma^{2/3}
  \approx 0.3048 \gamma^{2/3}$ and 
 two roots if $0< \mu <  0.3048 \gamma^{2/3}$.
 
 According to Eq.(\ref{pi9}), when $\gamma$ grows, the ``smaller" (stable)
 fixed point tends to $\epsilon= \frac{9}{16} \mu$;
 this reproduces Eq.(\ref{separa})  of subsection \ref{both_large}.

\begin{figure*}

\includegraphics[height = 2.0in, width = 0.45\linewidth]{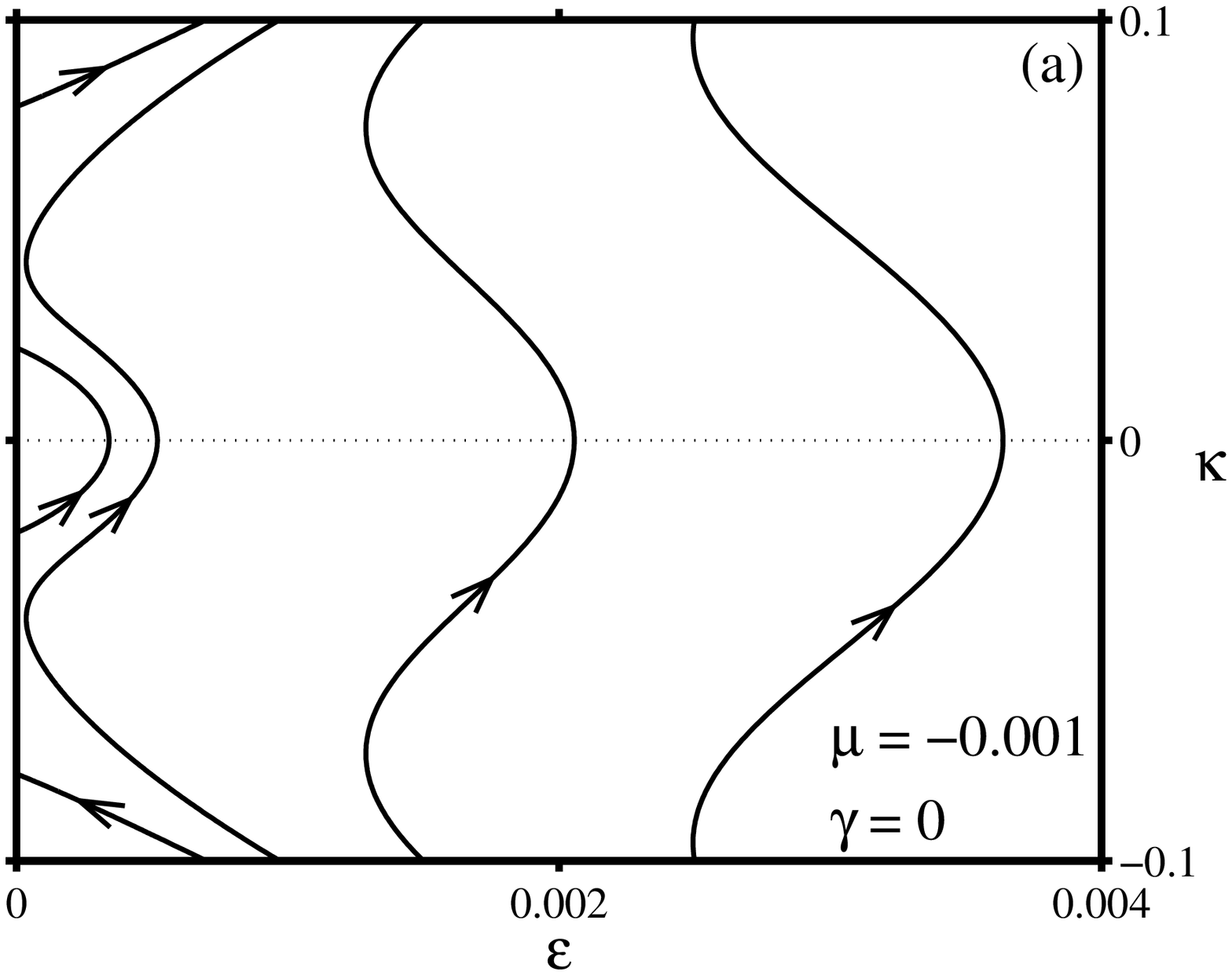}
\includegraphics[height = 2.0in, width = 0.45\linewidth]{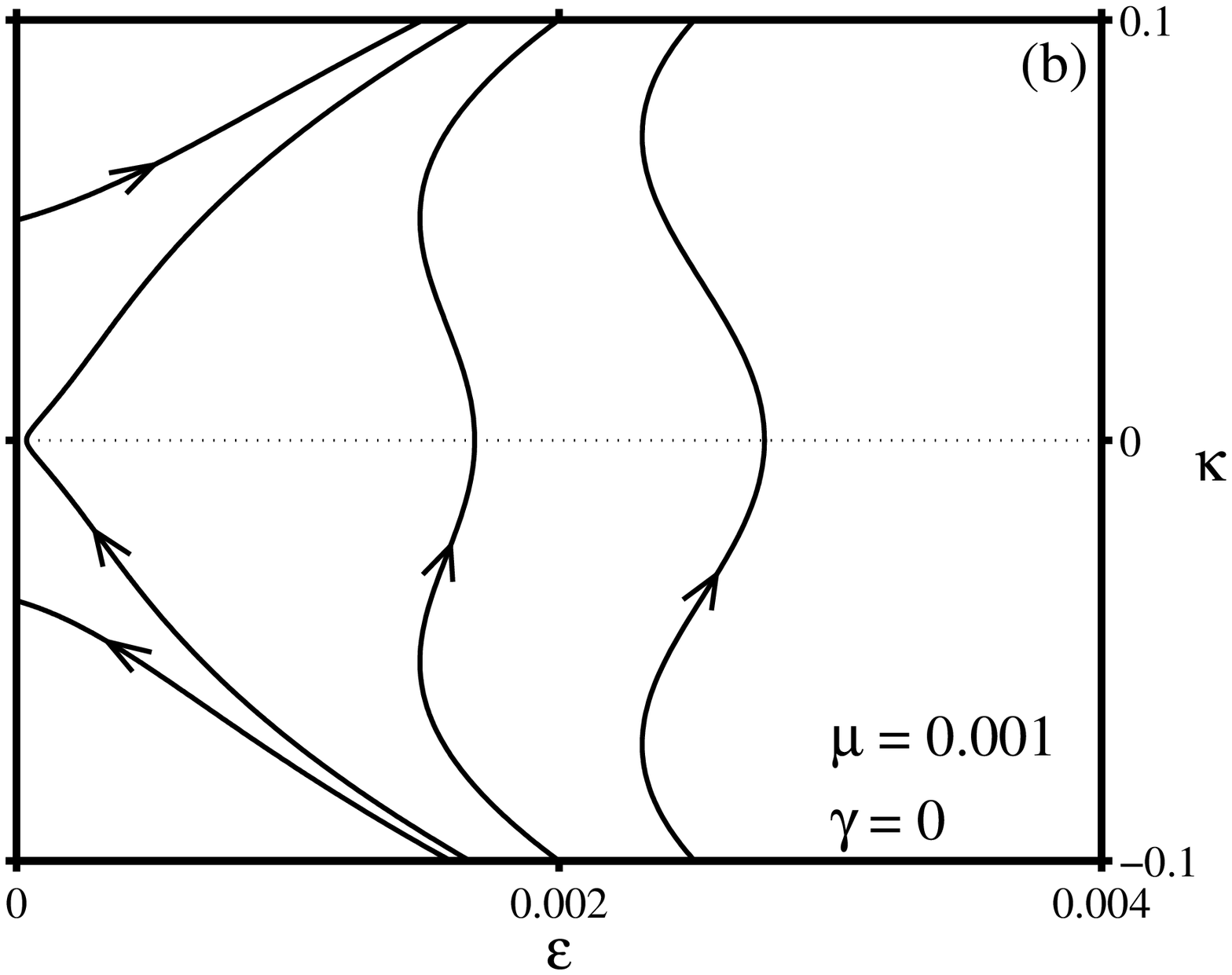}

\caption{\sf  The vector field (\ref{revvy})
for $\gamma=0$. (a) $A^2< \frac43$; (b) $A^2> \frac43$. }
\label{fig3}
\end{figure*}

These considerations translate into the following criterion
valid for small $\gamma$
and small  differences $A^2-\frac43$. 
[We are also assuming that the
imaginary parts of the perturbation of the walls are small
initially, i.e. 
$\kappa(0)$ is small.] 
The direction of forces between the walls depends on
their position relative to two stationary bound states,
a stable  (with the interwall separation denoted $2z_s$)
and an unstable one (with the  separation $2z_u$).
The unstable complex exists for all $A^2 < A_{sn}^2$
while the stable one only for $\frac43 < A^2 < A_{sn}^2$,
where
\begin{equation}
A_{sn}^2= \frac43 + 0.3048 \gamma^{2/3}.
\label{Ac_ga}
\end{equation}
In their region of coexistence ($\frac43 < A^2 < A_{sn}^2$),
the stable complex has a larger separation: $2z_s> 2z_u$. 
As $A^2 \to \frac43$, the distance $2z_s$
  grows to infinity.
  
When $A^2< \frac43$, two N\'eel walls  repel if their 
separation distance is greater 
than  $2 z_u$.
If the initial separation is smaller than $2 z_u$, the walls attract. 
It is natural to expect that the attraction should
result in  the annihilation of the walls, 
but this diagnosis is beyond the scope of the variational 
approximation and can only be established in direct
numerical simulations of the full  partial differential equation
(\ref{NLS2}) (see below).

Next, if  $\frac43 < A^2 < A_{sn}^2$, 
 the walls with separations larger than $2 z_u$
  will move
towards the stable equilibrium 
[from inside or outside depending on whether $z(0)<z_s$ or $z(0)>z_s$].
The walls at shorter distances, $2z(0)< 2z_u$, are attracted
to each other; they converge and, apparently, 
annihilate.

Finally, for $A^2> A_{sn}^2$, the walls are attracted to
  each other.  
  No complexes, stable or unstable,
   can be formed for these 
  large $A$.

Like in the large-$|\mu|$ situation, 
the undamped case ($\gamma=0$)
is exceptional in the small-$|\mu|$  region. 
In this case, the whole $\kappa$-axis is a line of fixed points. 
  When $\mu>0$, the phase portrait [Fig.\ref{fig3}(b)]
is similar to Fig.\ref{IBfig11}(b). The walls which are initially
at rest
[$\kappa(0)=0$], attract ($\epsilon \to \infty$ as $t \to \infty$).
As for the {\it moving\/} walls with large 
separations, their initial velocities are determined by $\kappa(0)$: 
\begin{equation}
{\dot z}(0) \approx
 -\frac{\kappa(0)}{\pi} \left[6 \mu + \frac{32}{9} \kappa^2(0) \right]. 
 \label{z_dot}
\end{equation}
If the walls are initially 
diverging sufficiently fast, they will continue to do so, with
\[ 
\kappa(t) \to \kappa_* <0; \quad
z(t) \to -\frac{\kappa_*}{\pi} \left(6 \mu + \frac{32}{9} \kappa_*^2
\right) t. 
\]

The portrait for $\mu<0$ is in
Fig.\ref{fig3}(a). Restricting our analysis to small 
$|\kappa(0)|$, we observe that walls with large separations 
repel and small separations attract. This is true
both for initially quiescent and slowly moving walls.

\subsection{Numerical study of the fixed points; 
all $\gamma$}
\label{numerical_FP}

As we have seen in section \ref{NeelVar},
  distant walls with $A^2> \frac43$ are attracted to each other.
Whether the attracting walls are going to
 form a stable bound state or collide and annihilate, 
 depends on two factors: (i) whether a stable and unstable  
 bound states exist for the corresponding  $\gamma$ and $A^2$,
 and if yes, then 
 (ii) how the initial interwall separation compares with $2 z_u$,
 the 
 interwall separation in the unstable complex.
 For small $\gamma$,  the region where the two complexes
 exist was found to be $\frac43< A^2< A^2_{sn}$, with 
 $A^2_{sn}$  as in Eq.(\ref{Ac_ga}). In order
 to classify the outcomes of the wall attraction, we need to
 demarcate the corresponding region for nonsmall $\gamma$ as well.
 
 The one-dimensional dynamical system (\ref{BSg})
 has only one, {\it stable\/}, fixed point.
  In order to describe the {\it unstable\/} complex
  (which is  bound tighter than the stable one, i.e.
  has a greater $\epsilon$), it is sufficient to
  retain the next, ${\cal O}(\epsilon^7)$, term in Eq.(\ref{310})
  when we substitute (\ref{parabola}) for $\kappa$.
  However, despite capturing both the stable and unstable 
  bound states, the system (\ref{parabola})-(\ref{310}) will not produce any 
  reasonably-accurate description of the  
   saddle-node bifurcation where 
  the two fixed points merge and disappear.
  One reason for this is that the term $32 \pi \gamma \epsilon^2 \kappa^2$ 
  in Eq.(\ref{31})
  which was dropped from Eq.(\ref{310}), becomes {\it larger\/}
  than $\epsilon^7$ when we substitute (\ref{parabola}) for
  $\kappa$. Another problem is that Eq.(\ref{parabola})
  keeps only the leading term  in the $\kappa(\epsilon)$ 
  expansion; retaining the next, ${\cal O}(\epsilon^5)$, term in
  $\kappa(\epsilon)$ produces another term larger than $\epsilon^7$ in
  Eq.(\ref{BSg}).

  Thus, in order to obtain an
  accurate variational description of the saddle-node 
  bifurcation,  we need to return to the original, nonreduced, vector
  field 
  (\ref{general}). Denoting $f(\epsilon, \kappa)$ and 
  $g(\epsilon, \kappa)$ the right-hand sides of (\ref{31}) and
  (\ref{30}), the two fixed points
  arise as intersections of the nullclines 
  \begin{equation}
  f(\epsilon, \kappa)=0, \quad g(\epsilon, \kappa)=0.
  \label{null_int}
  \end{equation}
  The nullclines have a common tangent
  when
  \begin{equation}
  \frac{f_\epsilon(\epsilon, \kappa)}{ f_\kappa(\epsilon, \kappa)}=
  \frac{g_\epsilon(\epsilon, \kappa)}{ g_\kappa(\epsilon, \kappa)}.
  \label{com_tan}
  \end{equation}
  Using a standard numerical routine to solve the system
  (\ref{null_int})-(\ref{com_tan}) 
  for the  vector of unknowns $(\epsilon, \kappa, A^2)$,
  one can find the saddle-node value $A_{sn}^2$ for each $\gamma$.
  
  The resulting curve $A^2=A^2_{sn}(\gamma)$ is shown in Fig.\ref{Edash2}
  below (the dotted line),
  where it is compared to the data from  the numerical continuation 
  of  the wall complexes as solutions to Eq.(\ref{NLS2}). 
  The significance of the curve is that it provides a borderline between
  the two dynamical scenarios:
  For $A^2 > A_{sn}^2$,
  two N\'eel walls attract each other and annihilate;
  for 
  $\frac43< A^2< A_{sn}^2$, the walls attract
  and annihilate if the initial separation $2z(0)< 2z_u$
  and form a stable bound state if $2z(0)> 2z_u$.
  
Note that the above numerical result is valid both for 
large {\it and\/} small $\gamma$. For small $\gamma$, 
our numerical curve $A^2_{sn}(\gamma)$ is reproduced by the
explicit formula (\ref{Ac_ga}).

We conclude this section by mentioning that the {\it unstable\/}
fixed point exists for all $A^2<\frac43$ for which there are 
N\'eel walls, i.e. for $1 < A^2 < \frac43$. This is a result
of the numerical study of the system (\ref{null_int})
with  $\gamma$ varying from $0$ to 
$1$ in steps of 0.01, and $A^2$ varying from  $1$
to $\frac43$ in steps of 0.001.
The reservation that we should make here, however, is that when 
$\gamma$ tends to 0 and, simultaneously, $A^2$ tends to 1, 
the coordinate $\kappa$
of the saddle point approaches a value of order 1.
This contradicts the assumptions we made in the derivation
of the system (\ref{general}) and so the
  fixed point in this parameter region cannot represent 
  any bound states of the walls.
 (The value of the $\kappa$-coordinate is reasonably small only
 when $\gamma$ is greater than 0.5 or when $|A^2-\frac43|$ is
smaller than 0.01.)
Consequently,  
the variational analysis cannot provide a trustworthy
description of the small-distance dynamics of the walls
in the $\gamma \approx 0$, $A^2 \approx 1$ region.
 The direct numerical simulations of the 
full  NLS equation (\ref{NLS2}) seem to be
the only reliable source of information here.

\section{Bound state of two dissipative N\'eel walls}
\label{DBS}

In the previous section we showed 
that two bound states of N\'eel walls,
a stable and an unstable one, exist in certain parts of the 
$(\gamma, h)$-plane. Here, we will reobtain 
these solutions numerically,  demarcate their regions of existence
and compare these to the corresponding variational results.

 \begin{figure}
 
\includegraphics[height = 2.5in, width = 0.5\linewidth]{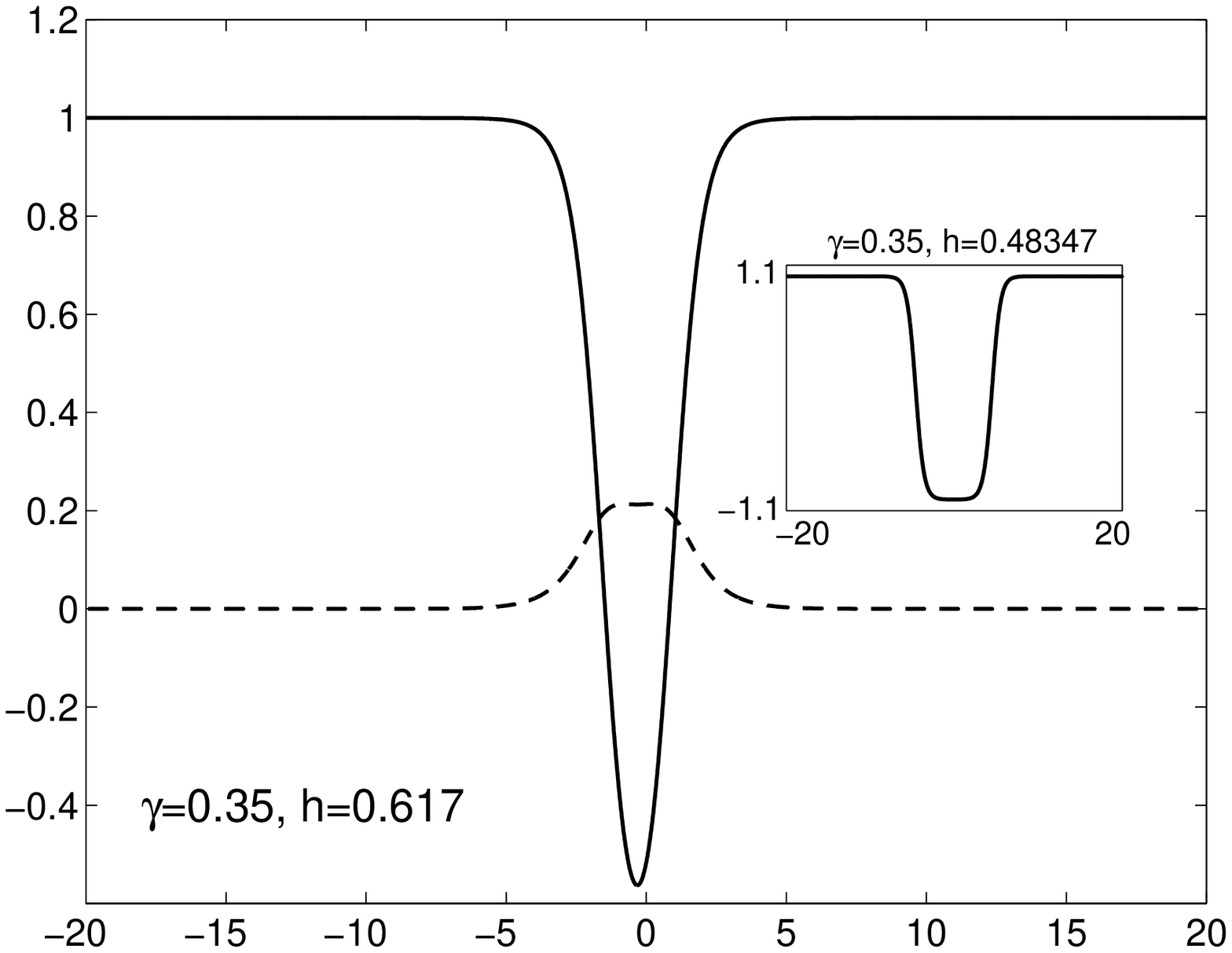}
 \caption{\sf
 A bubble with a small distance between the walls
 (a tightly bound complex) looks 
 like a single entity. (Note that the real part 
 does not even reach $-1$ in the core region.) The solid and the dashed
 lines
 stand for the real and imaginary part of $\psi$, respectively.
  For comparison, we 
 also show a bubble with a wide interwall separation 
 arising for $h^2$ close to $\gamma^2+ 1/9$ (inset). 
 Here only the real part is shown as ${\rm Im} \, \psi(x)=0$.} 
\label{bubble}
\end{figure}

 We will be using the term ``solitonic
 bubble" as a synonym for the wall complex. 
 Treating bound states as independent soliton-like entities
 is meant to emphasise the 
 significance of these stationary solutions
 as possible attractors in the phase space; 
 also, it should reflect their different, nontopological, nature.
 In fact, when the complex is tightly bound, it looks more 
 like a single entity than a wall doublet; see Fig.\ref{bubble}.
 
 \begin{figure}
\includegraphics[height = 2.5in, width = 0.5\linewidth]{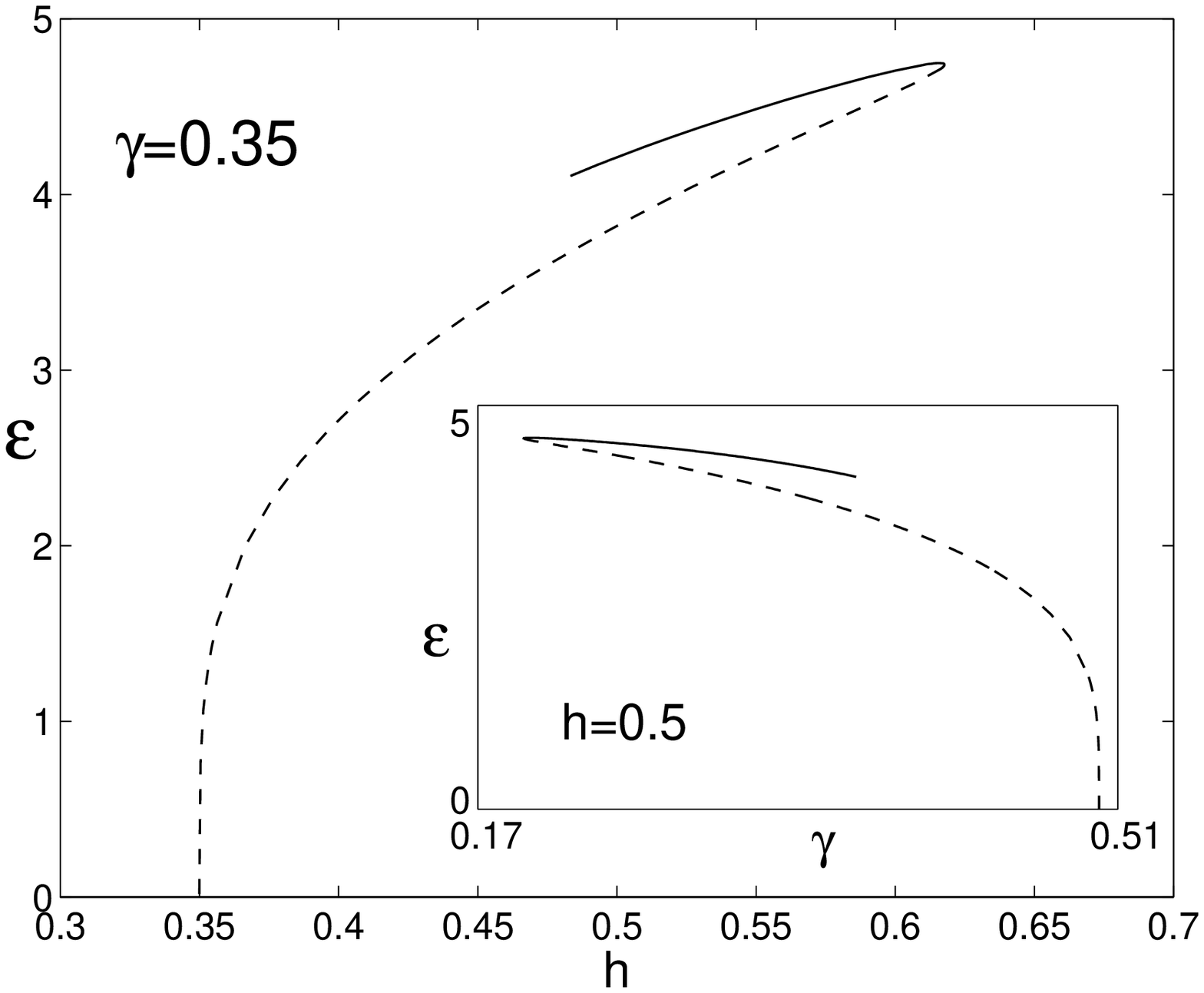}
 \caption{\sf   The energy of the dissipative bubble 
for the fixed $\gamma$  
(main panel) and 
 fixed $h$ (inset). The solid 
and dashed branches indicate stable and unstable solutions,
respectively. The stable branch  terminates at
 the point where the  distance between the walls 
 becomes infinite; this point is found to coincide with $A^2 =
\frac 43$. 
This termination scenario was predicted in
section \ref{NeelVar} using the variational method.} 
\label{Edash1}
\end{figure}
 
Our approach here  is based on the 
 numerical continuation
 (path-following)  of the stationary bubbles as solutions
 to the ordinary differential equation
\begin{equation}
 {\textstyle \frac {A^2}{2}} \psi_{xx} - A^2 |\psi|^2 \psi + \psi 
  + (A^2 - 1)\psi^* 
+ i \gamma(\psi - \psi^*)
= 0.  
\label{NLS_stat}
\end{equation}
The variational approximation (\ref{Ansatz}) with parameters found 
in subsection \ref{small_small} was used as an initial guess
for the numerical solution at small $\gamma$ and $h \approx \frac13$;
this, in turn, served as a starting point for our continuation.
We classified the stability of the resulting stationary complexes
as solutions of the full 
partial differential equation (\ref{NLS2}), by linearising about the 
solution and examining the spectrum of the linearised operator
numerically.

To present results of the continuation graphically, we use the energy 
functional 
\begin{equation}
E = \frac 12 \int \left\{  |\psi_x|^2 +  |\psi|^4 - 2\frac{ |\psi|^2  
+ h \, {\rm Re} (\psi^2) }{A^2} + 1 \right\} \, dx. 
\label{Energy}
\end{equation}
 Although the quantity defined in  Eq.(\ref{Energy}) 
 is not conserved for nonzero $\gamma$, 
the energy can be used as a bifurcation measure for stationary
solutions. 
[The field momentum (\ref{momentum})
 is not suitable for this purpose
  as it  satisfies  $\dot{P} = -2\gamma P$\  
 and so all stationary  solutions with $\gamma \neq 0$ have the same,
 zero,
 momentum.]

\begin{figure}
\includegraphics[height = 2.5in, width = 0.5\linewidth]{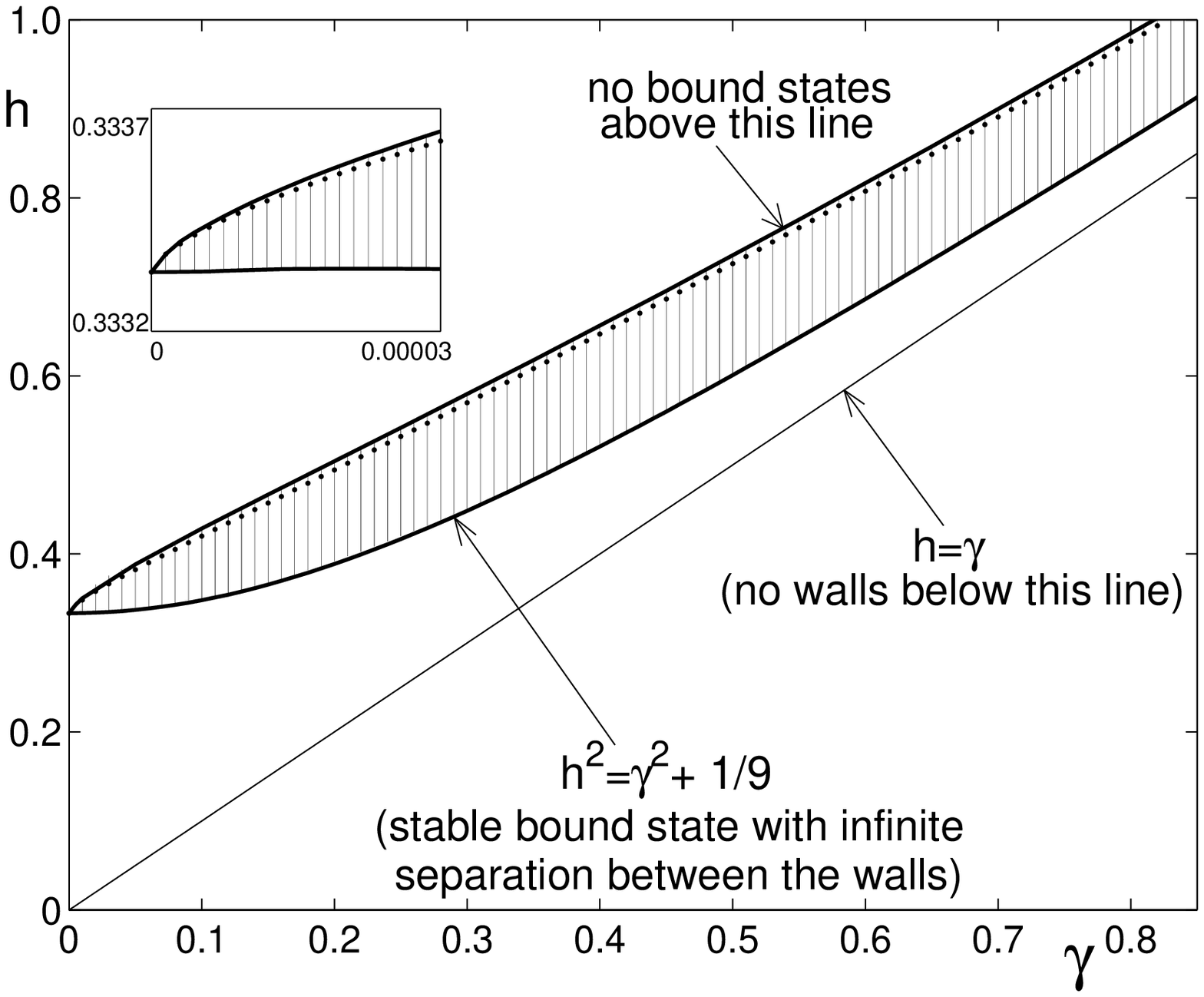}
\caption{\sf  The existence and stability diagram of the
bound states on the $(\gamma,h)$-plane (obtained
numerically.) The shaded corridor marks the region
of coexistence of the stable and unstable bubble. Its upper boundary 
corresponds to the turning point where the two bubbles merge;
there are no bound states above this line.
The dotted curve gives the variational approximation 
(\ref{null_int})-(\ref{com_tan})
to the line of turning points.
The lower boundary of the shaded area corresponds
to the stable bound state of two infinitely separated walls 
(an infinitely long stable bubble). 
This numerical curve is visually
indistinguishable from
 $h=\sqrt{\gamma^2+1/9}$. In the strip between $h=\sqrt{\gamma^2+1/9}$
and  $h=\gamma$, only the unstable bubble is found.
 Inset: a blow-up
of the small-$\gamma$ region.
Here, the dotted curve is plotted using an asymptotic formula 
(\ref{Ac_ga}). } 
\label{Edash2}
\end{figure}

A typical pair of bifurcation diagrams (corresponding
to the separate continuations in  $h$\ and $\gamma$) 
is presented in Fig.\ref{Edash1}.
[Fig.\ref{Edash1} corresponds to $\gamma=0.35$
(main frame) and $h=0.5$ (inset);  
the  continuations for other fixed values of $\gamma$ 
and $h$ produce curves of similar shape].  
Each diagram consists of two branches; 
the  entire branch with higher energy is found to be linearly stable.
The solutions on the lower branch are found to be unstable; the
instability is associated with 
 a (single) positive real eigenvalue. 
Numerical simulations of the full time-dependent
PDE (\ref{NLS2})
show that when perturbed, the unstable bubbles  
decay to the flat solution.

As we approach the
termination point
of the top branch, 
the separation of the walls becomes infinite.
  The termination point satisfies 
  $h \approx \sqrt{\gamma^2 +1/9}$ (i.e. $A^2 \approx \frac 43$) ---
 therefore, this solution
reproduces the stable bound state detected by the variational method
(sections \ref{both_large}, \ref{small_small}). 
The solution on the lower branch corresponds to
the unstable complex.
We note that this lower branch can be continued  
all the way to 
$h = \gamma$, in agreement with 
 variational results in section
 \ref{numerical_FP}.

The walls pull closer together as we move 
towards the turning point along the
upper branch (in the main frame of Fig. \ref{Edash1}).
As we continue away from the turning
point along the bottom branch, the distance between the walls
continues to decrease, reaches a minimum, and then starts increasing. 
This nonmonotonicity can be  explained in terms of the
fixed points of the dynamical system (\ref{revvy}).

The existence and stability of the
 bound states is summarised in Fig.\ref{Edash2}.
 This figure charts the $(\gamma,h)$-plane into regions of
 different type of interaction between the walls.
  In the shaded region and
 above it,  distant walls
 attract each other. The attraction may result in the creation 
 of a stable bound state at some finite separation; this
 happens in the shaded domain. 
 [The formation of the bound state
  is illustrated in Fig.\ref{SimulationPics2}(c) below.]
 Above the shaded corridor, two attracting N\'eel walls 
 are expected to collide 
 and annihilate.
 Finally, in the region between the shaded area and the bisector
 line $h=\gamma$, two  N\'eel walls at large separation
 repel.

 In Fig.\ref{Edash2}, we also display the variational 
 results for the saddle-node bifurcation curve $h_{sn}(\gamma)$.
 The dotted line in the main frame gives the result
 of solution of the system of three equations   
 (\ref{null_int})-(\ref{com_tan}) while the dotted
 line in the inset was plotted using Eq.(\ref{Ac_ga}).
 The variational results are seen to be in good
 agreement with the outcome of the numerical continuation.

 \section{Numerical simulations:
 N\'eel walls}
\label{Simulations}

\begin{figure}
 
\includegraphics[height = 2.0in, width = 0.5\linewidth]{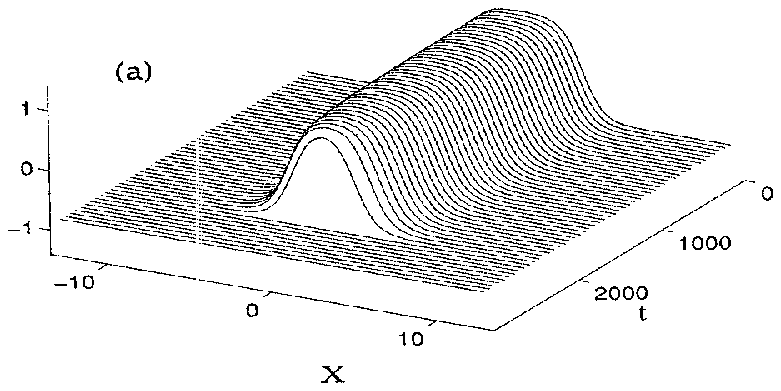}
\includegraphics[height = 2.0in, width = 0.5\linewidth]{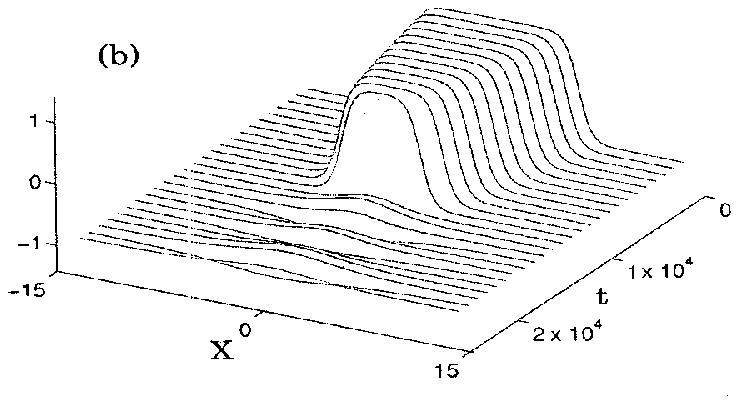}
\includegraphics[height = 2.0in, width = 0.5\linewidth]{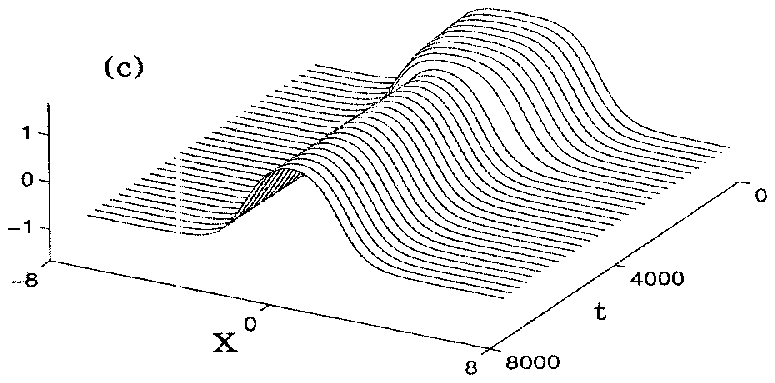}

\caption{\sf  
Attraction of two dissipative
N\'eel walls may result either in the annihilation of two walls
(a) (here $\gamma = 0.1$, $h = 0.5$) or formation of a bound state
(c)  (here $\gamma =0.1$, $h = 0.41$).
In the absence of  damping, 
the attraction results in the formation of a stationary breather (b)
(here $h = 0.5$).
Simulations reported in panels (a,b)
correspond to a point $(\gamma,h)$ lying above the
shaded region in Fig.\ref{Edash2} while 
those in panel (c) pertain to a point inside this region.
All panels show the real part of $\psi$, 
  multiplied by $-1$ 
for better visualization.
}
\label{SimulationPics2}
\end{figure}

Conclusions of the variational analysis were verified in 
direct numerical simulations of the NLS equation (\ref{NLS2}).
We used a split-step pseudospectral method
\cite{Herbst}, 
with $N = 512$\ Fourier 
modes on the interval $(-L/2, \, L/2)=(-40, \, 40)$.
The numerical scheme is stable  for the timesteps $\Delta t < 
\pi^{-1} (L/N)^2$; accordingly, we set 
$\Delta t= 2.5\times 10^{-3}$. The method imposes periodic boundary conditions. 
The 
initial condition had the form (\ref{Ansatz}) where  $\kappa_1$\ 
and $\kappa_2$\ 
were set equal to zero.
The results are shown in Figs.\ref{SimulationPics2}, 
where we
have ``zoomed in'' on the solitons neglecting the flanks of the
simulation interval.

To verify the diagram of Fig.\ref{Edash2}, we examined three sequences
of points on the $(\gamma, h)$-plane: one just above the shaded region,
one just below it, and one inside. 
The sequence just above the saddle-node line
consisted of ten points $\gamma=0.0, 0.1, ..., 0.9$ with $h=0.4+ 0.75 \gamma$.
For each pair of parameter values, we examined the initial separations
$2z(0)=3.0, 4.0$ and $5.0$.
The results  
were in agreement with the diagram of Fig.\ref{Edash2}:
all simulations 
 gave rise to the convergence 
 of the walls. 
 For $\gamma \neq 0$, this resulted in the annihilation of the walls
 [Fig.\ref{SimulationPics2}(a)] 
 while in the dissipation-free case,  the convergence of the walls
 ended in the formation
 of a nonpropagating breather [Fig.\ref{SimulationPics2}(b)].

The sequence just below the 
line $h=\sqrt{\gamma^2+1/9}$ consisted of   
$\gamma=0.0, 0.1,..., 0.9$ with $h=\sqrt{\gamma^2+0.09}$.
In the $\gamma=0$ case, the walls separated by
$2z(0)=4$ were observed to repel 
whereas those placed at a smaller 
distance $2z(0)=3$, 
attracted and formed a nonpropagating breather.
These results are in agreement with the 
variational analysis  [see Fig.\ref{fig3}(a)]. 
When $\gamma$ is nonzero, the initial separations
$2z(0)=3.0$ and $4.0$ gave rise to 
the repulsion of  the walls. In order to observe 
the attraction, we had to reduce the initial distance to
$2z(0)=1.4$ for $\gamma=0.1$;
to $2z(0)=0.8$ for $\gamma=0.2$ and to $2z(0)=0.1$ for $\gamma=0.4$.
These observations are consistent with the fact that 
$z_u$, the separation in the unstable
complex, becomes smaller as $\gamma$ grows.

 Finally, the sequence of points through the dashed
region in Fig.\ref{Edash2} included $\gamma=0.1,0.2, ..., 0.8$ with 
$h=\frac13 +0.784 \gamma$. For these values of $h$, the variational 
method predicts a tightly bound  stable complex
of two walls and indeed, in all eight 
cases the walls with initial separations $2z(0)=3.0, 4.0$ and $5.0$
moved towards each other and formed 
a stable bound state at a certain smaller distance
[Fig.\ref{SimulationPics2}(c)].

Thus the simulations of the N\'eel wall interactions
are in agreement with our expectations for distant-wall
dynamics summarised in Fig.\ref{Edash2}.

\section{Two Bloch walls of opposite chiralities}
\label{OppVar}

Proceeding to the analysis of Bloch walls, 
we remind the reader that 
the Bloch walls exist only for $\gamma = 0$\ and $h < \frac 13$; 
these will be
our assumptions in this section.
Here, we consider the interaction of two Bloch walls of 
opposite chiralities (i.e. with total field momentum  $P = 0$).
The corresponding lagrangian arises by setting
$\kappa_1=\kappa_2=\kappa$ and ${\cal B}=B$
in Eq.(\ref{LLL}). The resulting equations of motion
can be cast in the form of a hamiltonian system
\begin{equation}
 \dot{z}= \frac{1}{\pi \tau} \frac{\partial H}{\partial \kappa},
\quad
 \dot{\kappa}= -\frac{1}{\pi \tau} \frac{\partial H}{\partial z},
\label{H_eqs_B}
\end{equation}
where  $\tau(\epsilon, \kappa)$ is given by Eq.(\ref{tau}), 
and the Hamiltonian 
\begin{multline}
H(\epsilon,\kappa)= -\frac{2A^2}{3B}(\kappa^2-\kappa_0^2)^2 +
\frac{16h}{B} \epsilon \kappa^2 \\
+\frac{8}{3B} \epsilon^2 
[7-9h-2(9+5h) \kappa^2 +(15+11h) \kappa^4
-4 A^2 \kappa^6]
\\+16 A^2 \epsilon^2 z (2 \kappa^2 -\kappa_0^2- \kappa^4),
\label{Hamiltonian_BB}
\end{multline}    
with $\epsilon = e^{-2Bz}$ and $\kappa_0$  as in (\ref{kappa_0}).
 Trajectories of the system  (\ref{H_eqs_B}) 
 are simply level curves of the function $H(\epsilon,\kappa)$.

When the interwall separation is large, $\kappa$ should
be close to its unperturbed value: $\kappa \approx \kappa_0$
or $\kappa \approx - \kappa_0$, where $\kappa_0$ is, in general,  
$\mathcal{O}(1)$, but  becomes small  
 when $h$ is near $\frac 13$. 
In the derivation of the variational equations, 
we assumed that the walls are
well-separated and each wall is close to its unperturbed form. 
Therefore, only regions around the fixed points 
($0,\kappa_0$) and ($0,-\kappa_0$)
can be interpreted within the  PDE
(\ref{NLS2}) with full certainty. 
(Trajectories outside
those regions may also have infinite-dimensional counterparts
but this requires verification using direct numerical simulations
of the full PDE).
 Note that $\kappa_0  > 0$\ is the value of the imaginary part 
of the right-handed Bloch wall at its center
[see Eq.(\ref{Bloch})], 
and 
$-\kappa_0 < 0 $\ is the imaginary part of the left-handed Bloch wall 
at its center.  
Therefore, the point ($0,\kappa_0$) represents a configuration of the 
right-handed wall at $x =+\infty$\ and the left-handed antiwall at 
$x = -\infty$. 
The point ($0,-\kappa_0$) corresponds to the 
left-handed wall at
$x = +\infty$ and the right-handed antiwall at $x = -\infty$.

\begin{figure}

\includegraphics[ height = 2.0in, width =0.5\linewidth]{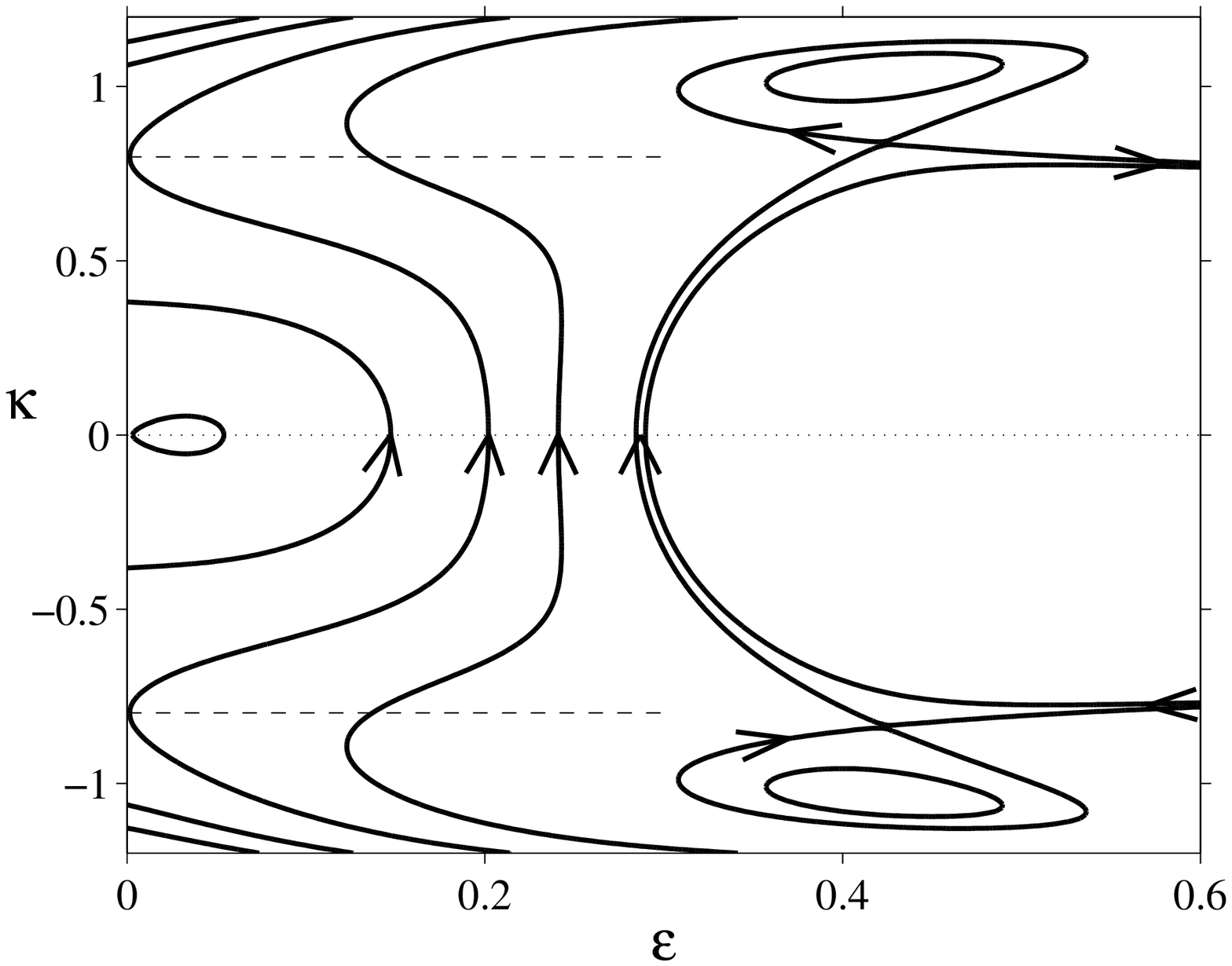}
\caption{\sf Trajectories of the 
system (\ref{H_eqs_B}) obtained as level curves of the Hamiltonian 
(\ref{Hamiltonian_BB}). 
(Here $h = 0.1$.) Two points at which the trajectory is tangent
to the $\kappa$-axis, correspond to stationary pairs of walls.
The dashed lines are $\kappa=\pm \kappa_0$;
the initial conditions lying on these lines pertain to
stationary walls: ${\dot z}(0)=0$. 
}
\label{Schweik}
\end{figure}

Assume, first, that $h$ is not close to $0$ or $\frac13$.
 When $\epsilon$ is small and $\kappa$ is close 
 to $\pm \kappa_0$ (which are not small), terms in the third and second
lines in (\ref{Hamiltonian_BB}) are much smaller
than the second term in the first line, and can be safely disregarded.
The resulting dynamical system
\[
\dot{\epsilon}=\frac{16}{3} A^2 \epsilon \kappa (\kappa^2-\kappa_0^2), 
\quad
\dot{\kappa} = 32 h \epsilon \kappa^2,
\]
does not have 
 fixed points with nonzero $\epsilon$ and $\kappa$;
on the other hand, the entire $\kappa$-axis is a line
of nonisolated fixed points. 
All these points describe pairs of walls
moving with constant velocities $\pm {\dot z}$, where
\[
\dot{z} = -\frac{8(1+h) }{3 \pi B} \kappa ( \kappa^2 -\kappa^2_0).
\]
Points on the positive-$\kappa$ semiaxis with
$\kappa < \kappa_0$ are stable and those with $\kappa>\kappa_0$ unstable.
Points on the negative-$\kappa$ semiaxis with
$\kappa<-\kappa_0$ are stable and those with $\kappa>-\kappa_0$ unstable.
(See Fig.\ref{Schweik}.) 
Consequently, walls with large initial separation 
[$\epsilon(0)$ near $0$] and $\dot{z}(0)>0$ diverge 
to infinities ($\epsilon \to 0$),
whereas distant walls with $\dot{z}(0) \leq 0$ converge.

As for the pairs of walls whose separation is not
very large, the situation is 
straightforward for the
walls which are not moving initially [$\dot{z}(0)= 0$];
 the corresponding initial conditions lie on the straight lines
$\kappa= \pm \kappa_0$ (dashed lines in Fig.\ref{Schweik}.)
These initial conditions give
rise to the convergence of the walls (i.e. $\epsilon$ grows as $t \to
\infty$).
Otherwise, the type of interaction depends on whether the point 
 $(\epsilon(0), \kappa(0))$\ lies  to the left
or to the right 
of the heteroclinic trajectory
[the trajectory which connects the point $(0,-\kappa_0)$
to $(0,\kappa_0)$ --- this trajectory is tangent to
the vertical axis in Fig.\ref{Schweik}.] 

\begin{figure}
\includegraphics[height = 2.0in, width = 0.5\linewidth]{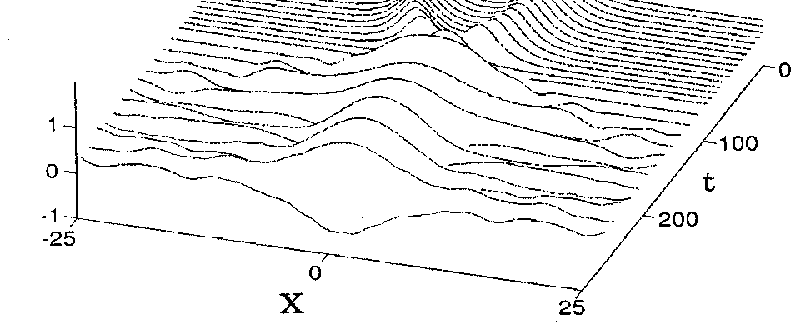}

\caption{\sf Attraction of Bloch walls with opposite chirality; this
attraction results in a long-lived breather. (In this plot, $h = 0.1$).
To highlight the  chiralities of 
the Bloch walls, the imaginary part of $\psi$ is shown. 
(We remind that in the case of the wall-antiwall
configuration that we are considering, 
 the same signs of the imaginary parts imply {\it opposite\/} 
chiralities of the walls.) 
}
\label{SimulationPics1}
\end{figure}

These variational conclusions are in agreement with 
direct numerical simulations of the full PDE (\ref{NLS2}).
(We employed the same numerical method as 
described in section \ref{Simulations}.)
The initial condition was chosen in the form  
of two Bloch walls of the 
opposite chirality which are initially at rest,
i.e. Eq.(\ref{Ansatz}) with $\kappa_1=\kappa_2=\kappa_0$.
 The initially quiescent walls have always
been observed to attract; see Fig.\ref{SimulationPics1}.

When  $h$ is close to  $\frac13$, $\kappa_0$ is
small; since $\kappa$ is assumed to be in the vicinity of $\pm \kappa_0$,
 we cannot neglect terms in the 
third and second lines in (\ref{Hamiltonian_BB})
in favour of the first line.
However, it is sufficient to keep just the leading term
(the term proportional to $7-9h$ in the second line).
A straightforward analysis reveals then that the
situation here is similar to the one considered above:
there are 
no fixed points with small nonzero $\epsilon$, while the
$\kappa$-axis is a line of nonisolated fixed points,
stable for $\kappa(\kappa^2-\kappa_0^2)<0$ and unstable otherwise.
Therefore the phase portrait for small $\epsilon$
and $\kappa$ is similar to the one above.

When $h$ is close to $0$, the last term in the first
line of  (\ref{Hamiltonian_BB}) becomes small and hence no 
terms in the third and second lines can be discarded. 
In this case, the dynamics can be analysed by plotting 
level curves of $H$ using standard software.
The same approach was adopted to study the vector field 
outside the small-$\epsilon$ region.

The resulting phase portrait is shown in Fig.\ref{Schweik}
for $h=0.1$; the portrait does not undergo any qualitative
transformations as $h$ is increased up to $\frac13$ or
decreased down to $0$. 
A striking feature of Fig.\ref{Schweik} is the existence of
three families of closed orbits: one centred on a point with
$\kappa \approx 1$; a mirror-reflected family centred 
on a point with $\kappa \approx -1$; and a family of 
closed trajectories centred on a point on the horizontal
axis. 

The closed trajectories centered on the fixed
point with $\kappa \approx 1$
 represent a family of 
breather solutions of the NLS (\ref{NLS})
(and so do their mirror-reflected twins). 
These breathers have been observed in our 
numerical simulations of the full partial differential equation
(\ref{NLS2}); see Fig.\ref{SimulationPics1}. They can
also be constructed perturbatively (see section \ref{Breathers} below). 
The nonlinear-center fixed point is 
sandwiched between two saddles, one below it (clearly visible
 in Fig.\ref{Schweik})
and one above. As $h$ is decreased, the 
two saddle points approach each other.
As a result, the periodic orbits are compressed 
in the vertical direction, and, 
as $h$ reaches zero, the family of closed trajectories 
degenerates into a segment of the straight line $\kappa=1$.

The  closed loops about a nonlinear-centre fixed point
with $\kappa=0$ represent another family of 
temporally-periodic solutions.
Since the value of $\kappa$ for a free-standing wall
(i.e. $\kappa_0$)
 is close to zero only when $h$ is close to $\frac13$, 
the family centred on $\kappa=0$ admits a reliable
interpretation only in this limit.   
When $h=0$, the  center point has the coordinate $\epsilon \approx
0.06$;
as $h$ grows to $\frac13$, the centre  moves towards the
origin so that the family of closed orbits shrinks
to a single point $\epsilon=0$, $\kappa=0$.
For $h$ close to $\frac13$,
i.e. in the region where these solutions can be interpreted 
in terms of the two-wall Ansatz (\ref{Ansatz}), the 
closed trajectories describe a pair of Bloch walls of opposite chiralities 
executing periodic oscillations of their separation 
(but remaining
far away from each other.)
These oscillatory ``doublets" have not yet been observed
in direct numerical simulations and their 
existence remains an open problem. (See the
concluding section 
for a possible explanation of the ``nonvisibility" of these
objects.)

\section{Two Bloch walls of like chiralities}
\label{LikeVar}

Finally, we examine the case of a pair of Bloch  walls 
(more precisely, a wall and an antiwall) 
of the same chirality.
 As in the previous section, 
we let $\gamma = 0$ and consider $h < \frac
13$. 
Assume, for definiteness, that we have a right-handed pair.
When at rest, the well-separated wall and antiwall 
have equal (nonzero) momenta $P_B= -\pi \kappa_0$
where  $\kappa_0$\ is as in (\ref{kappa_0}). When
the walls start moving as a result of their interaction, 
the total momentum remains the same and 
hence the individual field momenta of the two walls will
change by the same amount:  one will become $P_1=-\pi(\kappa_0+
q) $ 
and the other one $P_2=-\pi (\kappa_0-q)$. 
This means that one of the walls
will have the amplitude of its imaginary part increase 
by $q$  and the other one decrease by $q$,
and so we will not have a symmetric configuration 
of an equal-amplitude wall and antiwall any longer.
Therefore one may question the validity of our
assumption that the two walls remain equal distance away from the 
origin for $t>0$.
To see that the equal-distance Ansatz (\ref{Ansatz}) remains 
valid --- at least for not too late times --- 
 we note that the velocity-momentum curve $P(v)$ 
 of a single Bloch wall 
 has a 
 nonzero slope at the point $v=0$, $P=P_B$ 
 (see \cite{OurPaper}; the curve is also reproduced 
 in part II of the present series of papers \cite{BW2}). 
This means that the wall whose momentum has increased 
by $\Delta P=\pi q$, acquires the velocity $\Delta v$,
while the wall whose momentum has decreased 
by $\Delta P=\pi q$, starts moving with the velocity $-\Delta v$
where $\Delta v= (dP/d v)^{-1} \Delta P$. 
 Accordingly, if two walls are equal distance 
 from the origin initially, they will remain equally
far from the origin for all $t > 0$.
This symmetric arrangement will break down only
if $\Delta P$ is very large or if $(dP/dv)^{-1}$ is very small,
in which case one would have to take into account deviations
of the
shape of the $P(v)$ curve from a straight line.
The derivative $(dP/dv)^{-1}$ tends to zero only if $h \to \frac13$
and indeed, numerical simulations do reveal nonsymmetric motion of
like-chirality walls when $h$ is 
very close to $\frac13$  (see below), but outside this narrow region
the symmetric Ansatz (\ref{Ansatz}) remains valid.

\begin{figure*}
\includegraphics[height=2.0in,width =0.45\linewidth]{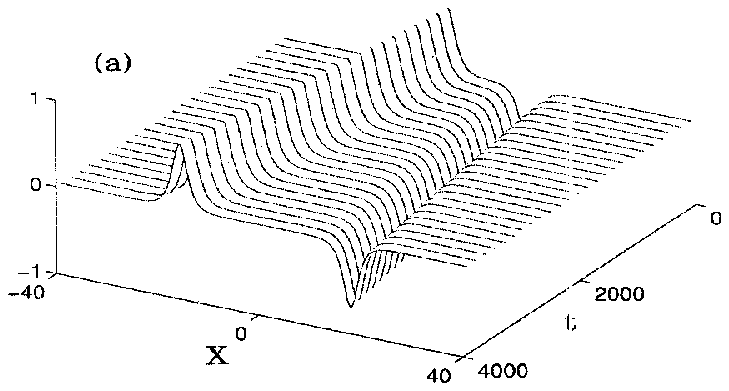}
\includegraphics[height=2.0in,width =0.45\linewidth]{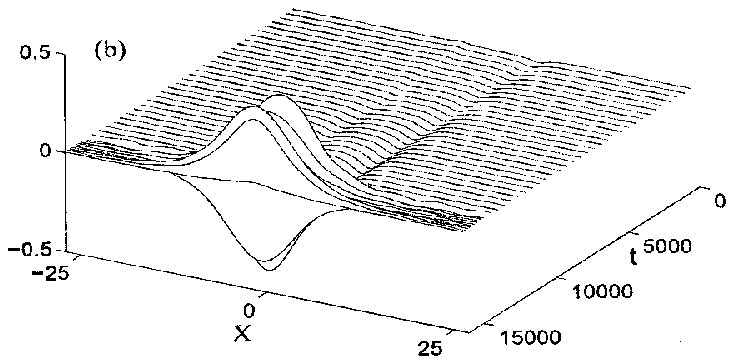}
\caption{\sf  Interaction of Bloch walls of  like chiralities. 
The imaginary part of $\psi$ is shown.
(a) Repulsion for generic $h$.  (Here $h = 0.15$).
(b) For $h$ close to
$\frac13$, the walls 
may attract and produce a breather on collision. 
(In this plot, $h = 0.333$).
Note that before the walls start converging, 
the wall on the left changes its chirality to the opposite.
(We remind that in the case of the wall-antiwall
configuration that we are considering, 
opposite signs of imaginary parts imply {\it like\/} 
chiralities of the walls.)}
\label{like}
\end{figure*}

Letting  
$\kappa_1 = \mp \kappa_0 + q$, $\kappa_2 = \pm \kappa_0 + q$\ 
and ${\cal B} = B$ in the Lagrangian (\ref{LLL}), we get
equations describing the dynamics of two 
like-chirality Bloch walls. These can be written as 
a hamiltonian system 
\begin{subequations}
\label{Hami_B}
\begin{equation}
 \dot{z}= \frac{1}{\pi \tau} \frac{\partial H}{\partial q},
\quad
 \dot{q}= -\frac{1}{\pi \tau} \frac{\partial H}{\partial z},
\label{H_eqs_N}
\end{equation}
where 
\[
\tau(\epsilon, q)=1+2  \epsilon
[1-14\epsilon -3(1-6\epsilon)\kappa_0^2 +(1+10 \epsilon)
q^2]>0
\]
and the Hamiltonian 
\begin{multline}
H(z,q)=
 -\frac{2A^2}{3B} (4 \kappa_0^2 +q^2)q^2
  -  \frac{16h}{B} \epsilon(\kappa_0^2-q^2) \\
+\frac{8}{3B} \epsilon^2
\left[ 7-9h - \frac{\kappa_0^2}{A^2} (7+38h+79h^2) \right. \\
\left. -\frac{4}{A^2} (3+12h+h^2) q^2 + (19-h)q^4 - 4A^2 q^6 \right] \\
+16 \epsilon^2 z [4h\kappa_0^2+4(1-h) q^2-A^2 q^4].
\label{H_like}
\end{multline}
\end{subequations}

\begin{figure*}
\includegraphics[height = 2.0in, width = 0.45\linewidth]{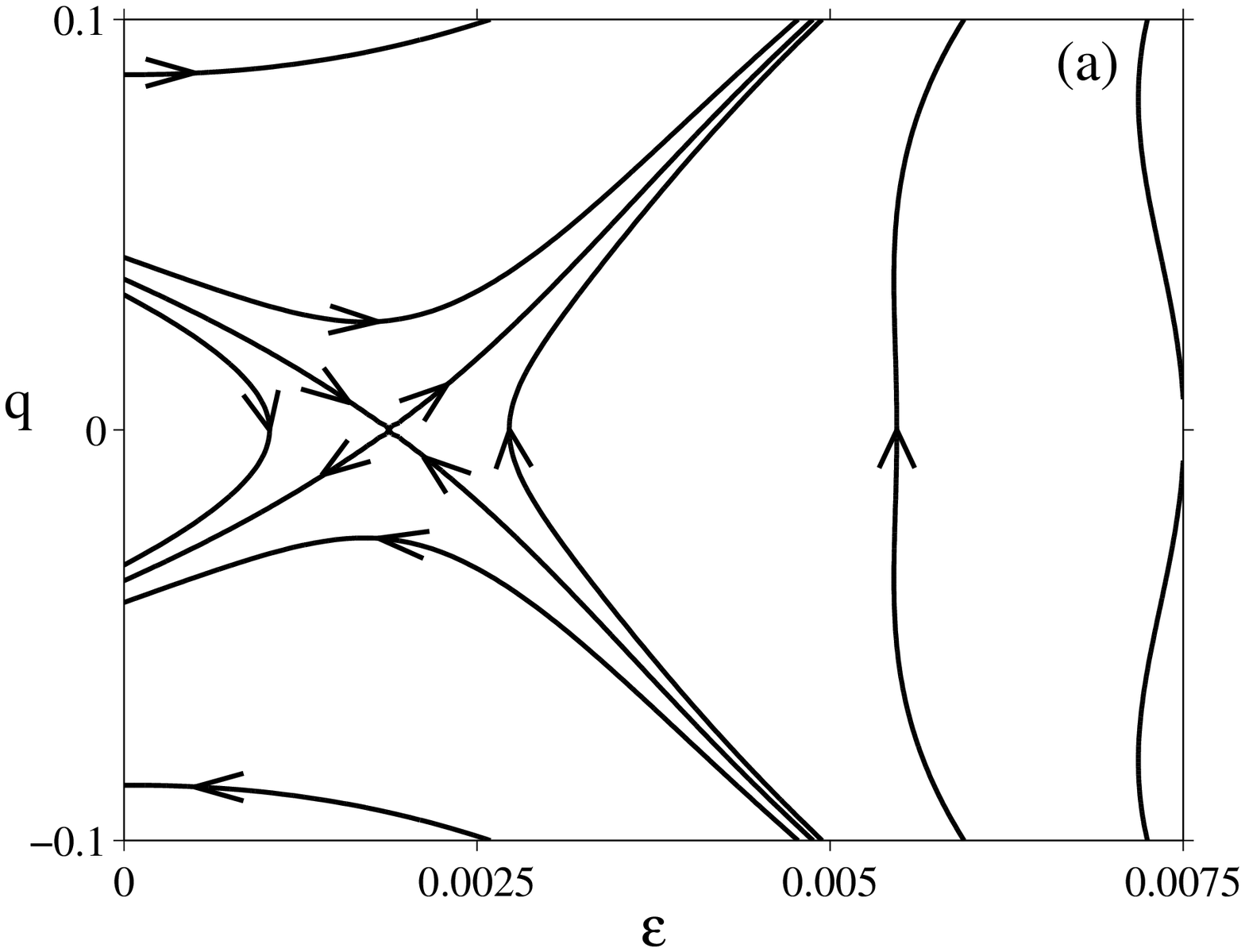}
\includegraphics[height = 2.0in, width = 0.45\linewidth]{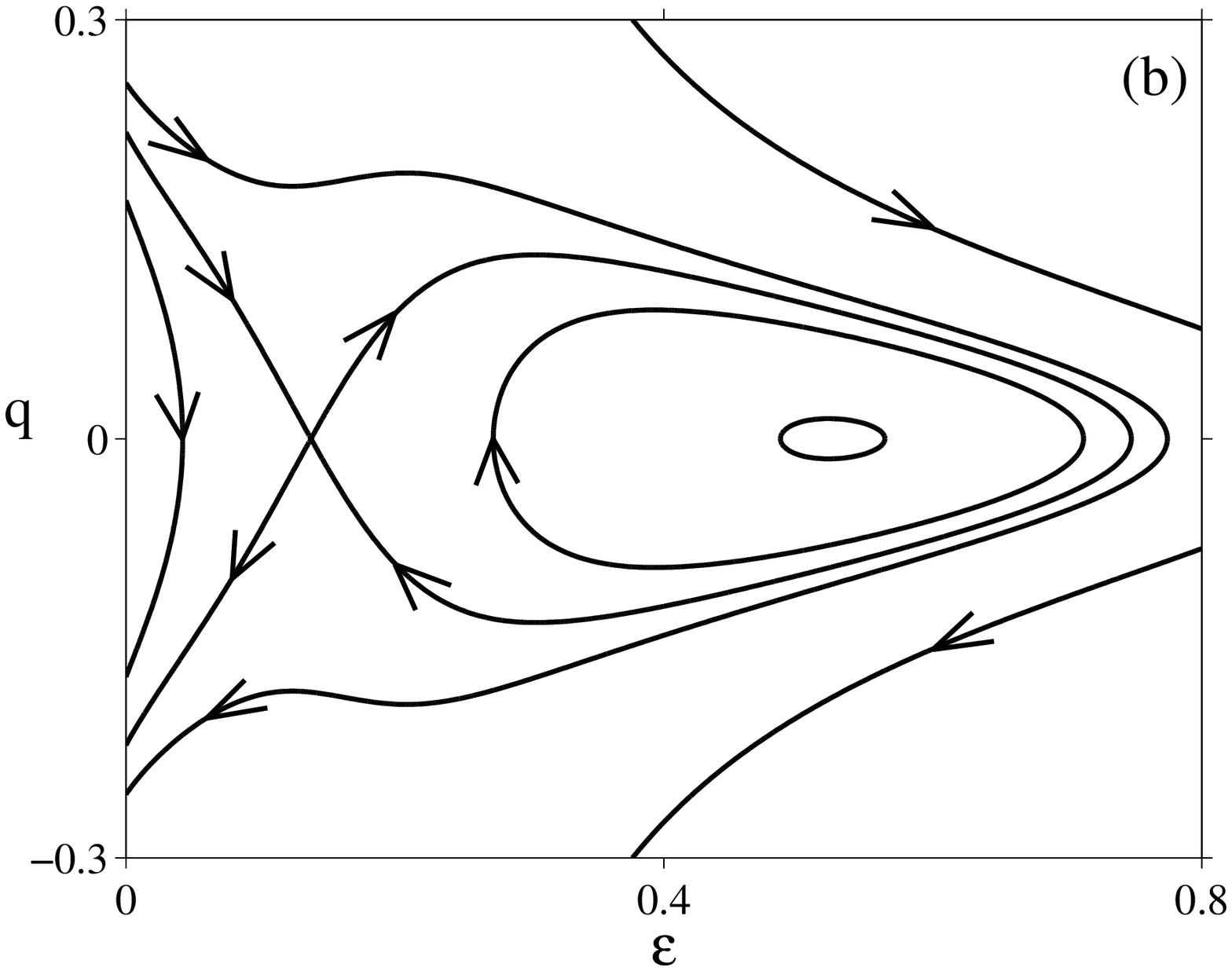}
\caption{\sf Trajectories of two Bloch walls with like chiralities on the 
($\epsilon, q$)-plane.   (a) $h$ close to $\frac13$ ($h=0.33$
in this plot). Note that 
the saddle point has small $\epsilon$ here; as a result, there is a region
of attraction within the range of validity of the variational
approximation. (b) Generic $h$ ($h = 0.2$ in this plot).
 The family of closed orbits
describe breather solutions of the NLS (\ref{NLS}).
Note that unlike in (a), the saddle point and the region of attraction 
arise at small interwall separations in (b).  
}
\label{Fig9}
\end{figure*}

Assume, first, that $h$ is not  close to $\frac13$,
so that $\kappa_0$ is not a small quantity. In this case
the Hamiltonian (\ref{H_like}) with small $\epsilon$ and $q$ 
is dominated just by
two terms, 
\begin{equation}
H(z,q) \approx
-\frac{8}{3B} A^2 \kappa_0^2 q^2 -
  \frac{16h}{B} \kappa_0^2 \epsilon.
  \label{Hzq}
\end{equation}
As in the previous section, the $q$-axis in this case is 
a line of nonisolated fixed points. 
 Since $\dot{z} \approx -(16A^2 \kappa_0^2/3 \pi B) q$, points with 
$q>0$ and $q<0$ represent pairs of converging and diverging
walls,
respectively.
The trajectories are parabolas $\epsilon=\epsilon_*- 
(A^2/6h) q^2$; they start at points on the positive-$q$ semiaxis
and flow into points on the negative-$q$ semiaxis.
This implies that the initially diverging 
walls continue to diverge whereas the
walls initially moving towards each other shall
stop at some finite separation, and then repel to infinities.
(A simple but practically important situation concerns 
walls which are initially at rest; these will obviously
start repelling straight away.)
The above conclusion applies, of course,
 only to slowly moving walls which
are sufficiently far from each other [i.e. $\epsilon(0)$ and $q(0)$ are 
small].

Fig.\ref{like}(a)  displays a typical result of direct numerical 
simulation of the interaction of two like-chirality Bloch
walls within Eq.(\ref{NLS2}). We tested several values
of $h$ and a variety of $z(0)$.
The initial condition was taken in the form (\ref{Ansatz})
with $\kappa_1=-\kappa_0$ and $\kappa_2=\kappa_0$.
In agreement with conclusions of the
variational analysis,  the initially quiescent walls were 
observed 
to repel for all $h$  not very close to $\frac13$.
(The case of $h$ close to $\frac13$ is discussed below.)

As $h$ tends to zero, the 
phase portrait does not undergo any qualitative changes.
However, the turning point of the parabola starting at a 
point $q_*$ on the $q$-axis, $\epsilon_*=(6h)^{-1/2} Aq_*$,
is shifted out of the small-$\epsilon$ region, i.e. out of the 
range of applicability of variational approximation.
Consequently, we can no longer expect the initially
converging walls to stop a large distance away from each
other and repel to infinities after that.
Instead, for $h \sim 0$ the colliding like-chirality Bloch 
walls should penetrate into the core of each other, and
the numerical simulations of the full PDE could be the only
way to study 
the outcome of this deep impact.

Next, let $h$ be near $\frac13$.
In this case
the Hamiltonian (\ref{H_like}) with small $\epsilon$ and $q$
reduces to
\[
H(z,q) \approx
-\frac89
(4 \kappa_0^2 q^2+q^4+ 6 \kappa_0^2 \epsilon -6 \epsilon q^2 -12
\epsilon^2).
\]
The dynamics with very small $\epsilon$ and $q$ is similar
to the one described by Eq.(\ref{Hzq}):
distant pairs of walls with $\dot{z}>0$ diverge 
while those which 
 are initially set to converge ($\dot{z}<0$)
  slow down, stop and then repel to
infinities [Fig.\ref{Fig9}(a)]. A new feature in 
the ($h \approx \frac13$)-case is a saddle point at 
$(\epsilon,q)=(\frac14
\kappa_0^2,0)$ and a region of attraction which
is separated from the region of repulsion by the 
stable manifold of the saddle [Fig.\ref{Fig9}(a)].
This fixed point
 was previously discovered 
 in a variational analysis 
of the (2:1)-resonantly forced Ginsburg-Landau equation 
\cite{MalNep}. 
Although it would be tempting to think that it represents 
a stationary complex of
two Bloch walls of like chiralities, 
 the actual interpretation 
 of the stationary point turns out to be somewhat different.  

Indeed, it was proved in \cite{OurPaper2} that in the 
absence of damping, two Bloch or two N\'eel walls cannot
form stationary complexes.
On the other hand, letting $\kappa_1=-\kappa_2=-\kappa_0$ and 
$e^{-2z}=\kappa_0^2/4$ with small $\kappa_0$, the trial function
(\ref{Ansatz}) can be written, approximately, as 
\begin{eqnarray*}
\mbox{Re} \, \psi = \frac{e^{2x} + e^{-2x} - e^{2z}}{e^{2x} +
 e^{-2x} + e^{2z}},\label{52} \\ 
\mbox{Im} \, \psi = \frac{4(e^x - e^{-x})}{e^{2x} + e^{-2x} + 
e^{2z}}\label{53}.
\end{eqnarray*}
This coincides with the expression for the 
Bloch-N\'eel bound state \{Eq.(11) of 
Ref.\cite{BW2}\} with $s = 0$\
and $\sigma = -1$, where we only need to send $B \rightarrow 1$\ 
and make the identification $e^{2\beta} = \frac 12 e^{2z}$. 
Therefore, the
``bound state" of two like-chirality Bloch walls produced 
by the variational analysis is, in fact, the 
Bloch-N\'eel complex with $s=0$. 
The reason why the Bloch-N\'eel complex could be mistaken for
a complex of two Bloch walls,  was simply because  
the Bloch and N\'eel walls become indistinguishable
as $h \to \frac13$.

The appearance of the region of attraction for $h$
close to $\frac13$ deserves a special comment.
The  attraction becomes possible due to the smallness of $\kappa_0$
in this limit.
Indeed, for sufficiently small $\kappa_0$, the momentum 
$\kappa_1=\mp \kappa_0 +q$ becomes close to $\kappa_2= \pm \kappa_0+q$
 ---
which is characteristic for two Bloch walls of 
{\it opposite\/} chirality. Thus, in the limit $h \to \frac13$, 
the walls may effectively change their chiralities.

It is instructive to consider the full phase portrait of the system
(\ref{Hami_B}); 
we produce it 
by plotting level curves of the 
Hamiltonian (\ref{H_like}) beyond the region of small $\epsilon$ and $\kappa$
[Fig.\ref{Fig9}(b)].
 A remarkable feature here is a family of closed orbits 
surrounding a nonlinear-center point.
The  center is born, along with the 
saddle point, in a saddle-center bifurcation as 
$h$ is increased past  $0.189$. The presence of periodic orbits
suggests that breathers may be formed in the collision of Bloch
walls of like
chirality. 

As $h \to \frac13$, the saddle point $(\frac14 \kappa_0^2, 0)$
moves closer to the origin and so the periodic orbits pass
near $\epsilon=0$. Accordingly, in this
limit the breathers should be accessible from 
initial conditions in the form of well-separated pairs of walls.
We note, however, that all closed orbits 
pass also through  a region of large $\epsilon$. 
[In fact, as  $h \to \frac13$, the coordinate
of the nonlinear center tends to infinity: $\epsilon_c \to
\exp(1/\kappa_0^2)$.]
Consequently, the formation of the breathers
requires accurate testing within the 
full PDE. In our numerical simulations 
with $h$ close to $\frac13$, we have indeed observed 
the attraction of walls with moderate separations
(e.g. with $\epsilon=0.005$ in the case of $h=0.33$ and 
$\epsilon=0.001$ in the case of $h=0.333$) followed by the
formation of a breather [Fig.\ref{like}(b)].
It is interesting to note that the attraction of
two like-chirality walls with $h \sim \frac13$ proceeds via
the chirality transmutation: first, one of the walls changes its
chirality to the opposite; after that, the two opposite-chirality walls
attract.

Finally, we need to mention an anomaly in
the interaction of two like-chirality walls which was observed in some 
of our simulations with $h$ near $\frac13$.
The walls which are initially at rest and equal
distance away from the origin, would start moving with
different velocities or even 
in the same direction --- violating our Ansatz 
(\ref{Ansatz}) based on the  
 assumption of a symmetric arrangement  for all $t$.
 This 
 asymmetric anomaly is observed in a wider range of $h$-values near $h=\frac13$
 than the 
 chirality transmutation 
 (we detected the former for $h$ as far from $\frac13$  as $h=0.30$)
 and 
 has a straightfoward explanation in terms
 of the $P(v)$ curve.  
 As $h \to \frac13$, the derivative $dP/dv$
 tends to infinity at the points where  the $S$-shaped $P(v)$-curve
 intersects the $P$-axis and so
 the equation $\Delta v=(d P/d v)^{-1} \Delta P$
 has to be amended by keeping the symmetry-breaking term in $(\Delta P)^2$.
 It is this term that causes the asymmetric motion of the 
 two walls.
 The anomalous interaction is typical for
 Bloch-N\'eel pairs  and can be interpreted as 
 the interaction of particles with the opposite mass sign \cite{BW2}.

 The fact that the motion of two walls has to
 be asymmetric becomes even more obvious if one 
 notices that the left 
 and right turning points
 of the $S$-curve approach $v=0$ as $h \to \frac13$. 
 As a result, the motion in the direction of the turning
 point becomes hampered. When $h$ is {\it extremely\/} close
 to $\frac13$, a new channel of interaction becomes available;
 namely,
 the wall which is repelled by its partner in the ``hard"
  direction, 
  may ``tunnel" through
 the $P(v)$-barrier  by means of
 the chirality transmutation, and the original symmetric arrangement remains
 undisturbed.

\section{Other localised attractors: nondissipative breathers}
\label{Breathers}

In the nondissipative situation
 ($\gamma = 0$), colliding walls  form a
moving or quiescent breather: 
a  temporally oscillating, spatially localised structure.
The breather was 
formed, for at least a short time,
in every undamped collision that we simulated. 
The formation of the breather is accompanied by 
the release of a large amount of
radiation; the numerical simulations reveal that the breather 
carries only about a quarter of the energy of the initial configuration, 
and so the
remaining three quarters must escape into the radiation field.
As a result, 
when the interval of simulation is too short,
 the interaction of the breather 
with the radiation re-entering the interval via the periodic boundaries 
is strong enough to 
destroy it  after just a few oscillations. On a large interval, 
however, the amplitude of the re-entering radiation is small 
 due to the dispersive
broadening, and the  breather persists indefinitely.

The breather can be easily constructed perturbatively. 
First of all, a harmonic solution $\delta \psi =  e^{i(\omega t - kx)}$\ of the NLS
(\ref{NLS2}) 
linearised about $\psi=1$
obeys the dispersion relation
\[ \omega^2
 = \left(1 + 2h + \frac{k^2}{2}\right)^2 - 1. \]
The $(k=0)$-harmonic oscillates with a nonzero frequency $\omega_0
= 2 \sqrt{h}A$; therefore, one may also expect to find a broad small-amplitude
breather oscillating with that frequency.   
We write
\begin{eqnarray}
\psi(x,t) = 
1 + \epsilon \psi_1(X_1,X_2, \dots, T_0, T_1, \dots) \nonumber \\ + 
\epsilon^2 \psi_2(X_1,X_2, \dots, T_0, T_1, \dots) + \dots,
\label{Breath1}
\end{eqnarray}
where  $X_n = \epsilon^n x$, $T_n = \epsilon^n t$\ and $\epsilon$\ is a
small parameter, and
substitute (\ref{Breath1}) in equation 
(\ref{NLS2}). Setting to zero
coefficients of $\epsilon^1$ and $\epsilon^2$, we 
find:
\begin{equation}
\psi_1 = u_1 + i v_1, \quad \psi_2 = u_2 + i v_2,
\label{Breath2}
\end{equation}
where 
\begin{equation}
\left(\begin{array}{c} u_1 \\ v_1 \end{array} \right) = 
\left(\begin{array}{c} 2h \\ i\omega_0 \end{array} \right) \frac{\phi}{2h}
e^{i\omega_0 T_0} + c.c.
\label{Breath3}
\end{equation}
and
\begin{equation}
\left(\begin{array}{c} u_2 \\ v_2 \end{array} \right) = 
\left(\begin{array}{c} 1+2h \\ i\omega_0 \end{array} \right) \frac{\phi^2}{2h} 
e^{2i\omega_0 T_0} - \left(\begin{array}{c} 1+4h \\ 
0 \end{array} \right) \frac{|\phi|^2}{2h} + c.c.
\label{Breath4}
\end{equation}
Here the amplitude $\phi$\ depends only on long scales
$X_1, X_2, ...$ and slow times $T_1, T_2, ...$, while 
$c.c.$\ indicates the complex conjugate. 
The solvability condition at the order $\epsilon^2$\ 
is $\partial \phi/ \partial T_1 = 0$, so
that $\phi$, in fact, is independent of $T_1$:
 $\phi = \phi(X_1,X_2,\dots, T_2, T_3, \dots)$.

At the order $\epsilon^3$, the solvability condition 
forces $\phi$\ to obey the attractive unperturbed nonlinear Schr\"odinger equation
\begin{equation}
-4i\frac{\sqrt{h}}{A} \frac{\partial\phi}{\partial T_2} + 
(1+2h)\frac{\partial^2\phi}{\partial X_1^2} 
 + 8(1+4h)|\phi|^2 \phi =0. 
\label{Breath5}
\end{equation}
The breather results if we choose $\phi$ in the form 
of the soliton:
\begin{eqnarray}
\phi = \phi_0 \, \mbox{sech}\left[2\sqrt{\frac{1+4h}{1+2h}} 
\phi_0 (X_1-VT_2)\right] \quad \quad \quad \nonumber \\ \times \exp\left[
-i\frac{A(1+4h)}{\sqrt{h}}
\phi_0^2 T_2 + V \frac{\sqrt{h}}{A(1+2h)} (2 X_1 - V T_2)\right], \nonumber \\
\label{Breath6}
\end{eqnarray}
where $\phi_0$\ and $V$\ are free parameters describing 
the amplitude and velocity of the soliton, 
respectively.
Substituting Eq.(\ref{Breath6})
into (\ref{Breath1})-(\ref{Breath3}) and defining 
the amplitude of the breather $a = \epsilon \phi_0$\ 
and its velocity $v = \epsilon V$, we obtain, finally,
\begin{widetext} 
\begin{subequations} 
\label{48} 
\begin{eqnarray}
\mbox{Re} \, \psi = 1 + 2a\, \mbox{sech}\left[2\sqrt{\frac{1+4h}{1+2h}}a(x- vt)\right] 
\cos \left[ \Omega t - v \frac{\sqrt{h}}{A(1+2h)}(2x - v t) 
\right] + \mathcal{O}(\epsilon^2), \\ \nonumber \\
\mbox{Im} \, \psi = -\frac{2 a A}{\sqrt{h}}\mbox{sech}
\left[2\sqrt{\frac{1+4h}{1+2h}}a(x-v t)\right] 
\sin \left[ \Omega t - v \frac{\sqrt{h}}{A(1+2h)}(2x - v t) 
\right] + \mathcal{O}(\epsilon^2),\,\,
\label{Breath7}
\end{eqnarray} 
\end{subequations}
\end{widetext}
where 
\begin{equation}
\Omega=  \omega_0 - \frac{A(1+4h)}{\sqrt{h}} a^2.
\label{Omega}
\end{equation}

We carried out numerical simulations of Eq.(\ref{NLS2}) with 
Eq.(\ref{48}) as an initial condition, with a variety of 
(small) values of $a$. 
(We confined our simulations to the case $v = 0$.) 
For all values of $h$\ and $a$\ we tried,
the breather persisted for the full length of the simulation 
(approximately $1000$\ periods of the breather's
oscillation), with virtually no change. 
The measured frequency of oscillation coincided with the asymptotic value 
(\ref{Omega})
up to
the third decimal place. 

Simulations were also performed with a gaussian initial condition 
\begin{equation}   
\psi = 1 + \alpha e^{-\beta x^2},
\label{Gaussian}  
\end{equation}
for a number of small complex values of $\alpha$
($|\alpha| \lesssim 0.5 $)  and 
positive $\beta$.
 All runs resulted in the formation of a breather, 
although in some cases [for $|\alpha| \gtrsim 0.3 $], the emerging breather 
would break into a pair of counterpropagating breathers. 
In those cases where 
the emerging breather was nonmoving, its shape and frequency 
were found to be in an excellent
agreement with the asymptotic formula (\ref{48}).

\section{Conclusions and Open Problems}
\label{Conclusions}

In this paper, we studied the interactions between the similar-type 
dark solitons 
of the nonlinear Schr\"odinger equation, i.e. Bloch-Bloch and
N\'eel-N\'eel interactions. Our approach was based on 
the variational approximation
(also known as the collective-coordinate, or 
variation-of-action, method).
The variational conclusions were verified 
using numerical continuation of solutions of the stationary
damped-driven NLS (\ref{NLS_stat}) and via direct numerical
simulations of the full partial differential equation
(\ref{NLS2}).

 \subsection{Conclusions}

In the dissipative situation 
($\gamma \neq 0$), the only available solitons
are the N\'eel walls.
When two N\'eel walls are very far apart, their interaction
is simple: the walls repel if $h^2< \gamma^2+ \frac19$
and attract if $h^2> \gamma^2 +\frac19$. The 
repelling walls diverge to infinities; as for  
the case of attraction, there are two possible scenarios.
In order to distinguish between the two, one is led
 to consider the situation 
where the walls are closer to each other [but still sufficiently
far apart for the 
 large-separation approximation (\ref{Ansatz}) to remain valid.]

At these shorter distances, the dynamics is influenced
by two bound states, a stable and an unstable one.
The stable bound state exists for $h$  between 
$\sqrt{\gamma^2+ 1/9}$ and a threshold
driving strength $h_{sn}$, and  the unstable one
exists for all $h<h_{sn}$. Here $h_{sn}^2=\gamma^2+(A_{sn}^2-1)^2$
where $A_{sn}^2(\gamma)$ is defined as a root of the system 
(\ref{null_int})-(\ref{com_tan}). For small $\gamma$, 
the curve
$h_{sn}(\gamma)$ can be described explicitly:
\begin{equation}
 h_{sn} = \frac13 + 0.3048 \gamma^{2/3}.
\label{hsn} 
\end{equation}
[This is Eq.(\ref{Ac_ga}) written
in terms of $h$ and $\gamma$.]
In their region of coexistence, the stable complex has a larger 
separation: $2z_s>2z_u$. 

When $h$ is smaller than $\sqrt{\gamma^2+1/9}$, 
the walls repel if their separation distance $2z(0)$
is greater than $2z_u$, the interwall separation in the unstable 
complex. If $2z(0)< 2z_u$, the walls  converge
and annihilate.
(This verdict does not extend to the region where 
{\it both\/} $h$ and $\gamma$ are small. In this region the 
variational analysis 
of the small-separation dynamics is inconclusive.)
When $\sqrt{\gamma^2+ 1/9}< h<h_{sn}$, pairs of N\'eel walls
with separations $2 z(0)$ larger than $2 z_u$ 
 evolve towards the stable bound state
while those with $2z(0)< 2 z_u$ converge
and annihilate. 
Finally, the walls with $h> h_{sn}$  converge
and annihilate
irrespectively of their initial separation.
These results pertain to walls at shorter distances (which 
are however sufficiently far apart for the variational 
approximation to remain valid).
In particular, they
 answer the question as 
to what finally happens to the two walls attracted 
from very large distances.

The nondissipative case requires a separate summary.
Here the walls can move at constant speeds 
and the interaction pattern 
becomes complicated by the presence of inertia.
When $h$ is greater than $\frac13$
(by small or large value),
the N\'eel walls attract
and converge  --- unless the initial condition corresponds
to walls having large and opposite velocities. In the 
latter case  
the attraction is unable to stop the diverging walls 
and they  escape to infinities.
On the other hand, when $h$ is smaller than $\frac13$, the walls
repel. The exception here is the case where $h$ is close to $\frac13$;
in this case  walls with very large separations
 repel whereas walls which are not so far from each other, attract.

In the dissipation-free case, the 
available dark solitons also include Bloch  walls.
The interaction between two Bloch walls depends on their
relative chiralities: 
two initially quiescent, oppositely-handed Bloch walls 
 attract while two quiescent 
walls with like chiralities 
placed at a large distance away from each other, repel.
The exception is the case of $h$ close to $\frac13$;
in this limit, two walls of like chirality repel at large distances
but exhibit anomalous interaction  or transmute into an opposite-chirality
pair and  
attract -- when placed closer to each other.

These conclusions can be extended to the case of the 
{\it moving\/} Bloch walls, where one 
just needs to take their inertia into account.
For example, two initially diverging 
oppositely-handed walls at large separation 
will continue to diverge despite the attraction
whereas two likely-handed 
walls which were initially moving against each other, will
continue to converge (until the 
repulsion stops them and sends away to infinities).

In addition to the interactions between well-separated walls, 
we  investigated products of their collision.
When the system is not damped, the collision of two 
walls results in a stationary or travelling breather.
We reconstructed the numerically found breather 
as an asymptotic series.

When $\gamma$ is nonzero, oscillations are damped and the
 only nontrivial
product of collision of two N\'eel walls
  is a stationary bubble ---
 the bound state of two N\'eel walls.
 Using the numerical continuation of solutions to the 
ODE (\ref{NLS_stat}), we demarcated the bubble's domain
of existence in the parameter space.
[This domain is, naturally, a subset of 
the part of the $(\gamma,h)$-plane in which remote walls attract].
The  numerical demarcation is 
in excellent agreement with  
the domain of existence obtained variationally.

\subsection{Open Problems}

\begin{figure}
\includegraphics[height = 2.5in, width = 0.5\linewidth]{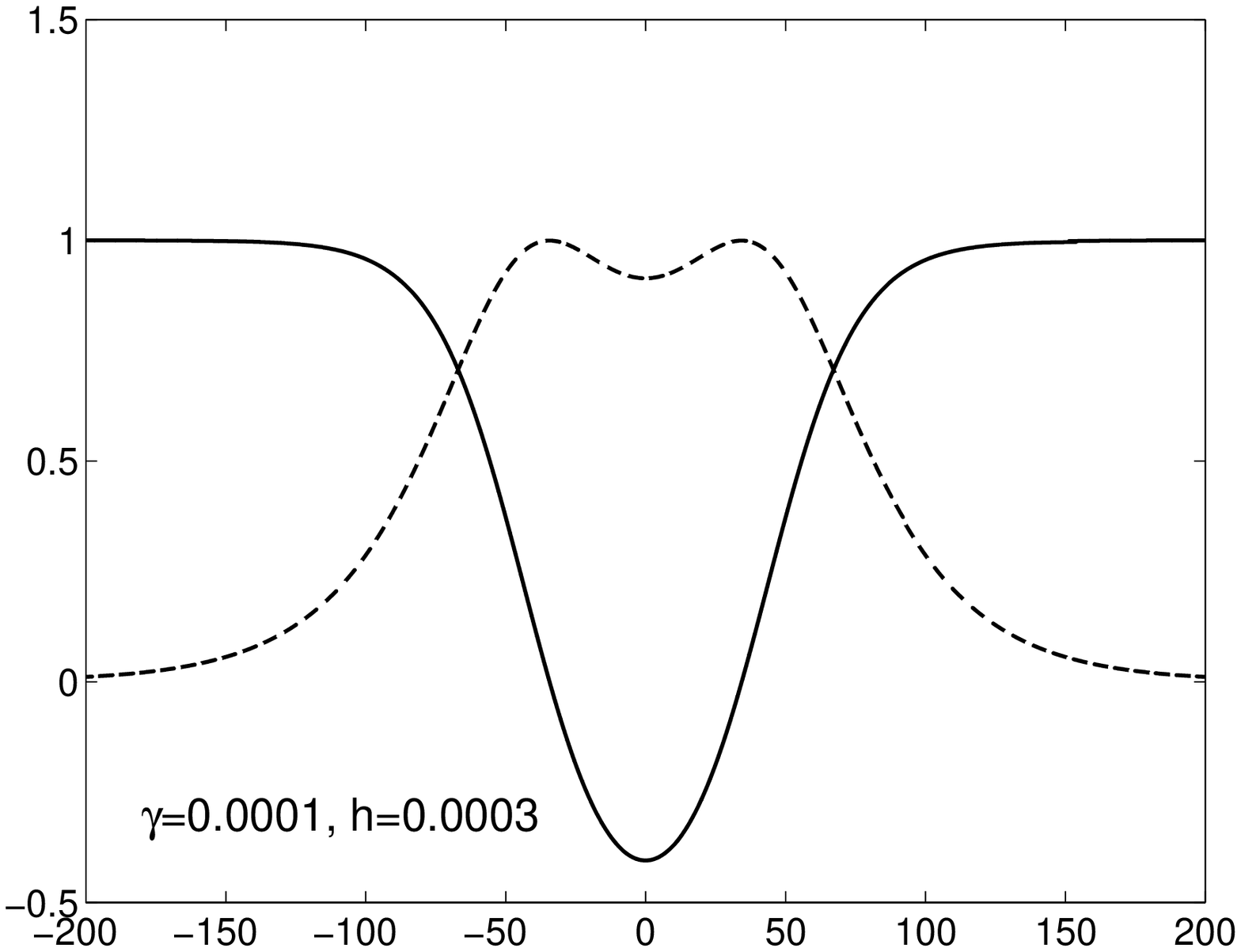}
\caption{\sf The tightly bound complex of two N\'eel walls with small 
$h$ and $\gamma$. The solid and dashed line
depict the real and imaginary part, respectively.
Note an enormous spatial extent of the bubble.}
\label{Weird_Bubble}
\end{figure}

1. We complete the paper by listing several open problems to which we plan
to return in future.
One interesting problem that merits further investigation, 
concerns the (unstable) bubble solution with small 
$h$ and $\gamma$. (This solution
is exemplified by Fig.\ref{Weird_Bubble}.)
Our numerics and the variational analysis in
section \ref{numerical_FP}  show that this solution {\it cannot\/} be
regarded as a complex
of two N\'eel walls, not even strongly overlapping ones.
Its asymptotic behaviour is closer to that of a Bloch wall; however,
Bloch walls do not exist in presence of damping.
 Furthermore, the bubble cannot
be continued to $\gamma=0$ and hence does not have a nondissipative
analogue in the form of a bound state of two Bloch, or a Bloch and
a N\'eel, walls. (This follows from the fact
that the N\'eel wall is the only solution in the $\gamma=0$ domain which
admits continuation  to nonzero $\gamma$ as it is the only
solution with zero momentum \cite{OurPaper}.)
Thus a question arises whether the bubble with small 
$h$ and $\gamma$ could
 be interpreted as a complex of two hypothetical
 ``dissipative Bloch walls" of opposite chiralities ---
 which do not exist individually but 
 can exist as a bound state due to the cancellation of
 their opposite momenta.  
 A similar momentum cancellation occurs
 in the attractive damped-driven NLS where it 
 allows complexes to travel with nonzero speeds
 despite the fact that their constituent solitons are 
 immobile \cite{SIAM}.

A related question concerns the short-distance N\'eel
wall dynamics with small $h$ and $\gamma$. This
problem is inaccessible to the variational
method, and the direct numerical simulations of the full
PDE seems to be the only appropriate line of attack here.

2. 
With regard to the Bloch walls, an
 open problem concerns the family
of closed orbits centred on $\kappa=0$ in
Fig.\ref{Schweik}. When $h$ is close to $\frac13$, these periodic orbits
can be interpreted within the full
partial differential equation  (\ref{NLS}),
as pairs of opposite-chirality Bloch walls executing periodic
oscillations of their separation. It is not impossible that 
these periodic solutions exist also for $h$ not so close to $\frac13$ 
[where the validity of the well-separated Ansatz (\ref{Ansatz})
becomes questionable.]
It remains to be understood why these orbits have not been observed in
our numerical simulations of Eq.(\ref{NLS}), with either $h$.
One possible explanation stems from the fact that 
 the nonlinear-center fixed point
enclosed by this family of closed orbits, corresponds to a {\it
maximum\/}
of the energy (\ref{VV}). Consequently, the nonlinear modes
not captured by our two-dimensional Ansatz (\ref{Ansatz}) ---
such as radiations --- should make the point $(\epsilon, \kappa)$ 
slide downhill towards orbits with larger radii, until
it crosses through the homoclinic trajectory and ends
up at $\epsilon=0$. [It is pertinent to note here that the 
centre points enclosed by the other two families of closed 
orbits in Fig.\ref{Schweik} are {\it minima\/} of the energy (\ref{VV});
this is consistent with the stability of the corresponding breather
solutions
observed in numerical simulations in Section \ref{Simulations}.] 

3. Next, it would be interesting to simulate collisions
of the like-chirality Bloch walls 
with $h \sim 0$. This case is not amenable to the 
variational analysis due to the large extent (``width") of
the walls.  

4. It would also be interesting to gain a deeper theoretical 
insight into the anomalous interaction of the like-chirality Bloch walls
arising for $h$ close to $\frac13$. 
The variational approach remains a proper tool here;
one should only 
generalise the Ansatz (\ref{Ansatz}) by
allowing nonsymmetric configurations of the walls:
\begin{eqnarray*}
  \varphi_1(x,t)= \tanh{[{\cal B} (x+z_1)]} 
  - i\kappa_1 \mbox{sech} [{\cal B} (x+z_1)],
  \label{Convenience1} \\
  \varphi_2(x,t)=\tanh{[{\cal B} (x-z_2)]} 
+ i\kappa_2 \mbox{sech} [{\cal B} (x-z_2)]. 
\end{eqnarray*}
Here $z_1, z_2$ and $\kappa_1, \kappa_2$ are 
unrelated pairs of variables.

5. Finally, one more future challenge is the analysis of the interaction 
of two breathers and their synchronisation to a 
common frequency.

\acknowledgments

It is a pleasure to thank Boris Malomed for useful references.
One of the authors (IB) thanks Dr Reinhard Richter
and Prof Ingo Rehberg for their hospitality at
the University of Bayreuth where this project was completed.
IB is a Harry Oppenheimer
Fellow;  also supported by the NRF of South Africa under grant 2053723. 
SW was supported by the NRF
of South Africa. EZ was supported by the Russian Foundation for Basic
Research under grant 06-01-00228.

\end{document}